\documentclass[twocolumn]{aa} % for a paper on 2 column  
\usepackage[update,prepend]{epstopdf}
\usepackage{txfonts}
\usepackage{graphicx}           % For eps figures, newer & more powerfull
\usepackage{mathrsfs}           % Calligraphic style
\usepackage{amssymb}            % Mathematical symbols
\usepackage[usenames,dvipsnames]{color}  % For color text: \color command
\usepackage{array}              % lines in tables
\usepackage{longtable}          % Split tables across multiple pages
\usepackage{natbib}             % For citations
\bibpunct{(}{)}{;}{a}{}{,}      % 

\newcommand{\bcAp}{bc\textsubscript{Ap}} 
\newcommand{\bcA}{bc\textsubscript{A}} 
\newcommand{\dVCAp}{dVC\textsubscript{Ap}} 
\newcommand{\dVCA}{dVC\textsubscript{A}} 

\newcommand{\eq}[1]{Eq.~(\ref{eq:#1})}
\newcommand{\Eq}[1]{Equation~(\ref{eq:#1})}
\newcommand{\eqs}[2]{Eqs.~(\ref{eq:#1},\ \ref{eq:#2})}

\newcommand{\eqss}[2]{Eqs.~(\ref{eq:#1} - \ref{eq:#2})}

\newcommand{\appx}[1]{Appendix~\ref{s:#1}}
\newcommand{\sect}[1]{Sect.~\ref{s:#1}}
\newcommand{\Sect}[1]{Section~\ref{s:#1}}
\newcommand{\tab}[1]{Table~\ref{t:#1}}

\newcommand{\fig}[1]{Fig.~\ref{fig:#1}}
\newcommand{\Fig}[1]{Figure~\ref{fig:#1}}

\newcommand{\BE}{\begin{equation}}
\newcommand{\EE}{\end{equation}}
\newcommand{\BA}{\arraycolsep=1.0pt\begin{eqnarray}}
\newcommand{\EA}{\end{eqnarray}}
\newcommand{\BI}{\begin{itemize}}
\newcommand{\EI}{\end{itemize}}

\newcommand{\Nabla}{\vec{\nabla}}

\newcommand{\rmd}{{\rm d}}

\newcommand{\dSa}{\rmd \vec{S}^a}
\newcommand{\dSb}{\rmd \vec{S}^b}
\newcommand{\dS}{\rmd \vec{S}}
\newcommand{\ds}{\rmd  {S}}
\newcommand{\dV}{\, \rmd \mathcal{V}}
\newcommand{\dVa}{\, \rmd \mathcal{V}^{\rm a}}

\newcommand{\surf}{{\partial \mathcal{V}}}
\newcommand{\surfa}{ {\partial \mathcal{V}^{\rm a}} }
\newcommand{\surfb}{ {\partial \mathcal{V}^{\rm b}} }
\newcommand{\surfS}{\Sigma}
\newcommand{\surfSa}{{\Sigma^a}}
\newcommand{\surfSb}{{\Sigma^b}}
\DeclareRobustCommand{\mSigma}{\text{\reflectbox{$\Sigma$}}}
\newcommand{\surfZ}{\mSigma}
\newcommand{\surfZa}{{\surfZ^a}}
\newcommand{\surfZb}{{\surfZ^b}}

\newcommand{\vol}{\mathcal{V}}
\newcommand{\vola}{\mathcal{V}^{\rm a}}
\newcommand{\volb}{\mathcal{V}^{\rm b}}

\newcommand{\ints}{\int_{\surf}}
\newcommand{\intv}{\int_{\vol}}
\newcommand{\intsa}{\int_{\surfa}}
\newcommand{\intsb}{\int_{\surfb}}
\newcommand{\intsS}{\int_{\surfS}}
\newcommand{\intsSa}{\int_{{\surfS}^a}}
\newcommand{\intsSb}{\int_{{\surfS}^b}}
\newcommand{\intza}{\int_{\surfZa}}
\newcommand{\intzb}{\int_{\surfZb}}

\newcommand{\intva}{\int_{\vola}}
\newcommand{\intvb}{\int_{\volb}}

\newcommand{\divA}{\Nabla \cdot \vA}

\newcommand{\divBp}{\Nabla \cdot \vBp}

\newcommand{\phia}{\phi^{\rm a}}
\newcommand{\phib}{\phi^{\rm b}}
\newcommand{\chia}{\chi^{\rm a}}
\newcommand{\chib}{\chi^{\rm b}}

\newcommand{\vA}{\vec{A}}
\newcommand{\vAa}{\vA^{\rm a}}
\newcommand{\vAb}{\vA^{\rm b}}
\newcommand{\vAab}{\vA^{\rm ab}}
\newcommand{\vApa}{\vA_{\rm p}^{\rm a}}
\newcommand{\vApb}{\vA_{\rm p}^{\rm b}}
\newcommand{\vApab}{\vA_{\rm p}^{\rm ab}}

\newcommand{\Az}{A_{\rm z}}

\newcommand{\vAp}{\vA_{\rm p}}

\newcommand{\Bp}{B_{\rm p}}
\newcommand{\vB}{\vec{B}}

\newcommand{\vBpa}{\vB_{\rm p}^{\rm a}}
\newcommand{\vBpb}{\vB_{\rm p}^{\rm b}}
\newcommand{\vBpab}{\vBp^{\rm ab}}
\newcommand{\vBp}{\vB_{\rm p}}
\newcommand{\vBj}{\vB_{\rm j}}

\newcommand{\vn}{\vec{n}}

\newcommand{\vx}{\vec{x}}

\newcommand{\hatn}{\hat{\vn}}
\newcommand{\hatna}{\hat{\vn}^{\rm a}}
\newcommand{\hatnb}{\hat{\vn}^{\rm b}}

\newcommand{\EB}{\mathscr{E}}
\newcommand{\HH}{\mathscr{H}}
\newcommand{\dHH}{\delta\mathscr{H}}
\newcommand{\dHHp}{\delta\mathscr{H}_{\rm p}}
\newcommand{\dHHmix}{\delta\mathscr{H}_{\rm mix}}

\newcommand{\Hv}{H}
\newcommand{\dHv}{\delta\Hv}

\newcommand{\Hmix}{\mathscr{H}_{\rm mix}}

\newcommand{\DHp}{\delta \HH_{\rm p} }

\usepackage[normalem]{ulem}     %Provides the following commands:
\newcommand{\eg}{\textit{e.g.},}
\newcommand{\ie}{\textit{i.e.},}

\newcommand{\strtable}{\renewcommand{\arraystretch}{1.2}} %Stretch vertical space in tables

\begin{document}
\title{
Additivity of relative magnetic helicity in finite volumes
}
% %  \subtitle{}
\titlerunning{Additivity of helicity}
\author{
Gherardo Valori\inst{1}
\and
Pascal D\'emoulin \inst{2}
\and
Etienne Pariat\inst{2}
\and
Anthony Yeates \inst{3}
\and
Kostas Moraitis \inst{2}
\and 
Luis Linan \inst{2}
}
\institute{
        $^{1}$  University College London, Mullard Space Science Laboratory, Holmbury St. Mary, Dorking, Surrey, RH5 6NT, U.K.   \email{g.valori@ucl.ac.uk}
        \\
        $^{2}$ LESIA, Observatoire de Paris, Universit\'e PSL, CNRS, Sorbonne Universit\'e, Universit\'e de Paris, 5 place Jules Janssen, 92195 Meudon, France
        \\
        $^{3}$ Department of Mathematical Sciences, Durham University, Durham, DH1 3LE, UK
          }
\authorrunning{Valori G. et al.}
\date{Received ***; accepted ***}

   \abstract
% context heading (optional)
{
Relative magnetic helicity is conserved by magneto-hydrodynamic evolution even in the presence of moderate resistivity.
For that reason, it is often invoked as the most relevant constraint to the dynamical evolution of plasmas in complex systems, such as solar and stellar dynamos, photospheric flux emergence, solar eruptions, and relaxation processes in laboratory plasmas.
However, such studies often indirectly imply that relative magnetic helicity in a given spatial domain can be algebraically split into the helicity contributions of the composing subvolumes, \ie{} that it is an additive quantity. 
A limited number of very specific applications have shown that this is not the case.
}
% aims heading (mandatory)
{
Progress in understanding the non-additivity of relative magnetic helicity  requires removal of restrictive assumptions in  favour of a general formalism that can be used both in theoretical investigations as well as in numerical applications. 
}
% methods heading (mandatory)
{
We derive the analytical gauge-invariant expression for the partition of relative magnetic helicity between contiguous finite-volumes, without any assumptions on either the shape of the volumes and interface, or the employed gauge.
}
% results heading (mandatory)
{
The non-additivity of relative magnetic helicity in finite volumes is proven in the most general, gauge-invariant formalism, and verified numerically. 
More restrictive assumptions are adopted to derive known specific approximations, yielding a unified view of the additivity issue. 
As an example, the case of a flux rope embedded in a potential field shows that the non-additivity term in the partition equation is, in general, non-negligible.
}
% conclusions heading (optional) 
{
The non-additivity of relative magnetic helicity can potentially be a serious impediment to the application of relative helicity conservation as a constraint to the complex dynamics of magnetized plasmas.
The relative helicity partition formula can be applied to numerical simulations to precisely quantify the effect of non-additivity on global helicity budgets of complex physical processes.
}

   \maketitle
\section{Introduction}\label{s:intro}
%Helicity: definition and properties, 
Magnetic helicity is a general measure of the complexity of magnetic fields that concisely expresses the amount of twist, writhe, and mutual winding of field lines in a given configuration \citep{Elsasser1956,Berger1999}. 
When applied to magnetized plasmas, the concept of  helicity acquires the very special role of an integral of motion.
Intuitively, the reason for that is that, as Alfv\'en's theorem demonstrates, in ideal magneto-hydrodynamics the topology of the magnetic field cannot be changed  by plasma motions. 
As ideal evolution cannot change the field topology, and therefore the field lines' entanglement, magnetic helicity is conserved in dissipationless (ideal) plasmas \citep{Woltjer1958b}, and nearly conserved in mildly collisional ones \citep{Matthaeus1982b,Berger1984}.
From a different perspective, the inverse cascade that characterizes magnetic helicity \citep[see, \eg][]{Frisch1975,Alexakis2006,Mueller2013a}, as opposed to the direct cascade of magnetic energy towards the small dissipative scales, is often invoked as the underlying paradigm behind the appearance of large-scale magnetic fields \citep[see, \eg ][]{Antiochos2013}.
Taken together, the conservation and inverse-cascade properties give magnetic helicity a unique power to describe the evolution of magnetized plasma.

%Recent progress and applications: towards an helicity budget
The concept of magnetic helicity is very general, and the wide applicability of magneto-hydrodynamics makes helicity a cross-disciplinary tool.    
In the solar context, for instance, it was applied to topics such as dynamos \citep{Brandenburg2005a}, reconnection \citep{DelSordo2010}, fluxes of helicity through the photosphere \citep{Pariat2005,Demoulin2009,Schuck2019}, the distribution of helicity in the corona \citep{Yeates2016b}, the initiation of coronal mass ejections \citep{Pariat2017b,Thalmann2019a} and their link to interplanetary coronal mass ejection \citep{Nakwacki2011,Temmer2017b}, just to name a few examples.
All these type of studies are related to each other by treating different aspects of the generation and evolution of the solar magnetic field that are constrained by the conservation of magnetic helicity.   

%Natural plasmas, finite volumes
Magnetic helicity is expressed as the volume integral of the magnetic field and its vector potential (see \eq{H} below), and is therefore gauge-dependent, unless the considered volume is bounded by a magnetic flux surface. 
Such a requirement is generally not satisfied by natural plasmas, nor in numerical simulations. 
In order to overcome this limitation, \cite{Berger1984a} and \cite{Finn1985} introduced the concept of \textit{relative} magnetic helicity, where the helicity in an arbitrarily shaped  finite volume is computed with respect to a reference field that has specific properties at the boundary.
In this way, the values obtained by the volume integral are made independent from the details used in the vector potential computation, \ie\ are gauge-invariant.

%H as constraint, and assumption of additivity
Insofar as the different processes involved can be described by magneto-hydrodynamics, the helicity of the field generated in the interior of the Sun must be conserved during the buoyant phase, through its rearrangement during the emergence through the photospheric layer forming long-lived coronal structures that finally erupt \citep[see \eg ][]{Priest2016}, to its propagation through interplanetary space \citep[\eg\ ][]{Demoulin2002,Green2002,Nindos2003a,Thalmann2019a}. 
In principle, a budget of (relative) magnetic helicity can be built that accounts for the transformation of the magnetic field from the interior of the Sun up to transient perturbations of interplanetary coronal mass ejections \citep[\eg{} ][]{Berger2000b,Demoulin2016a}. 
To exploit such a remarkable property requires the quantitative separation and comparison of, \eg{} the helicity emerged in the corona from the helicity left under the photosphere, or the helicity ejected as a coronal mass ejection  (and probed at the spacecraft position) from that left behind on the Sun. 

%Non additivity, previous results on this topic
Numerical simulations of different degrees of realism are available for all those processes.
However, there is a principal difficulty in separating the helicity into subvolume contributions: a limited number of very specific applications \citep{Berger1984a,Longcope2008a} have shown that  the sum of the relative helicity in two contiguous subvolumes is not simply equal to the helicity of the total volume. 
In this sense, relative magnetic helicity is not an algebraically additive quantity.
This is a serious impediment to the exploitation of the conservation principle in building global budgets of relative magnetic helicity.
In addition, the discussion of the additivity issue by \cite{Berger1984a} applies to volumes that are either unbounded or bounded by a flux surface, which are conditions that are not normally satisfied in numerical simulations.
Similarly, \cite{Longcope2008a} proposes an extension of the formalism in \cite{Berger1984a} that is intended to be applied to the finite volumes of numerical simulations, but still basically considers a coronal volume that is bounded above by a flux surface.
Finally, the choice of gauge made in \cite{Berger1984a} and \cite{Longcope2008a} is only one of the possibilities for discussing the non-additivity property, which may not be always available to specific applications where the gauge choice is limited by other factors (\eg{} by numerical precision). 

%Focus of this article
The main goal of this work is to derive general equations for the additivity of the relative magnetic helicity in finite volumes, and to provide a gauge-invariant expression for the non-additive terms.
In particular, the additivity problem is here formulated as a partition problem between two subvolumes that are contiguous and share a common boundary as, \eg\ in the flux emergence process where helicity is transferred from a sub-photospheric volume to the coronal volume.
The generality of our treatment is such that, on the one hand, it allows us identify the reason for the non-additivity in a general way.
On the other hand, our method  can be easily adapted to different geometries and gauge choices, allowing for straightforward applications to numerical simulations.

%Structure of the article
In \sect{Hpart} we discuss the nature of the additivity problem; we then derive the general partition equation without any assumptions on either the shape of the volumes and interface, or the employed gauge.
\Sect{gauges} gives a brief overview of how the partition equation is modified by the choice of commonly used gauges, which are then applied in \sect{models} to derive known expressions for the partition formula that are used in the literature. 
A numerical application to a solution of the force-free equation is used in  \sect{test} to verify the accuracy of the partition  formula and to test the importance of the non-additive term with respect to the helicity of the field.
Finally, in \sect{conclusions} we summarize our results, discuss their implication for the definition of the relative magnetic helicity, and propose a number of applications of our formalism.

\section{Partition of helicity between two volumes}\label{s:Hpart}

%%%%%%%%%%
\subsection{General definitions} \label{s:General_def}
As usual, for a magnetic field $\vB$ and its associated vector potential $\vA$, we define the gauge-dependent magnetic helicity $\HH$ in a volume $\vol$ as
\BE
\HH(\vB,\vol) = \intv  \vA \cdot \vB \, \dV \, ,
\label{eq:H}
\EE
and the gauge-invariant relative magnetic helicity \citep{Finn1985} as
\BE
 \Hv(\vB,\vol) =\intv \left ( \vA + \vAp \right ) \cdot  \left (  \vB -\vBp \right ) \, \dV \, ,  
 \label{eq:Hv}
\EE 
where the field $\vBp$ of vector potential  $\vAp$ satisfies 
\BE
\left . \hatn \cdot \vBp \right |_\surf =\left . \hatn \cdot \vB \right |_\surf
\label{eq:Bp_bc}
\EE
on the boundary $\surf$, with $\hatn$ the external normal to $\surf$. 
By construction, $\surf$ is a flux surface for the field $\vBj=\vB-\vBp$.
Any reference field that satisfies \eq{Bp_bc}, and the solenoidal condition $\divBp =0$, ensures the gauge-invariance of \eq{Hv}. 
In this work we assume that all magnetic fields satisfy the solenoidal condition exactly. 
A discussion of the consequences of the violation of the solenoidal condition in numerical computation of helicity and energy can be found in \cite{Valori2016} and \cite{Valori2013}, respectively. 

In principle, any field that satisfies \eq{Bp_bc} can be used as reference field, with \eq{Hv} defining the helcity relative to the chosen reference field. 
We adopt the common choice of a potential field as reference field $\vBp$.  
In this case, $\vBp$ is written in function of the scalar potential $\phi$ as $\vBp=\Nabla \phi$, where $\phi$ satisfies the Laplace equation 
$\Delta\ \phi =0$ in $\vol$ with the Neumann boundary condition
\BE
\left . \hatn \cdot \Nabla\phi  \right |_\surf = \left . \hatn \cdot \vB \right |_\surf   \, , \label{eq:Nabla_phi_bc} \\
\EE
such that the gauge-invariance requirement \eq{Bp_bc} is satisfied.
\Eq{Nabla_phi_bc} uniquely defines $\vBp$ in $\vol$, and defines $\phi$ in $\vol$ up to an additive constant.
The potential field defined by \eq{Nabla_phi_bc} has the minimal energy for the given distribution of the normal component of the field on the boundary, $\left .\hatn \cdot \vB \right |_\surf$, see, \eg{} \cite{Valori2013}. 
Therefore, this customary choice is not only convenient in its simplicity, but carries also a deeper physical meaning of the potential field being the ``ground state'' for a given distribution of field on the boundary, especially within the magneto-hydrodynamical framework, see \eg\ \cite{Schuck2019}.
Moreover, if $\vBp$ is the potential field in $\vol$, then $\vBj$ is the part of the magnetic field that is related to the presence of currents in $\vol$, sometimes referred to as the current-carrying part of the field. 
We adopt the potential field as defined above as reference field in the remainder of the article, but we explicitly discuss the consequences of this choice on our results, when relevant.

The relative magnetic helicity, \eq{Hv}, can be recast as the sum of three non-gauge-invariant terms: 
\BE
\Hv (\vB,\vol) = \HH(\vB,\vol) - \HH(\vBp,\vol) + \Hmix(\vB,\vBp,\vol) \, , \label{eq:nongi_decomp}
\EE
\ie\ as the difference between the magnetic helicity of $\vB$ and $\vBp$ plus a ``mixed term'' defined as 
\BA
 \Hmix(\vB,\vBp,\vol) &=& \intv \left (\vAp \cdot \vB - \vA \cdot \vBp \right) \ \dV \nonumber \\
                      &=& \ints \left ( \vA \times \vAp \right) \cdot \dS \ ,  \label{eq:HmixS}
\EA
where $\dS=\hatn\ \ds $ is the oriented infinitesimal surface element on $\surf$.

%----------------------------------------------------------------------------------------------
 \begin{figure}[t!]
  \centering
  \includegraphics[width=\columnwidth]{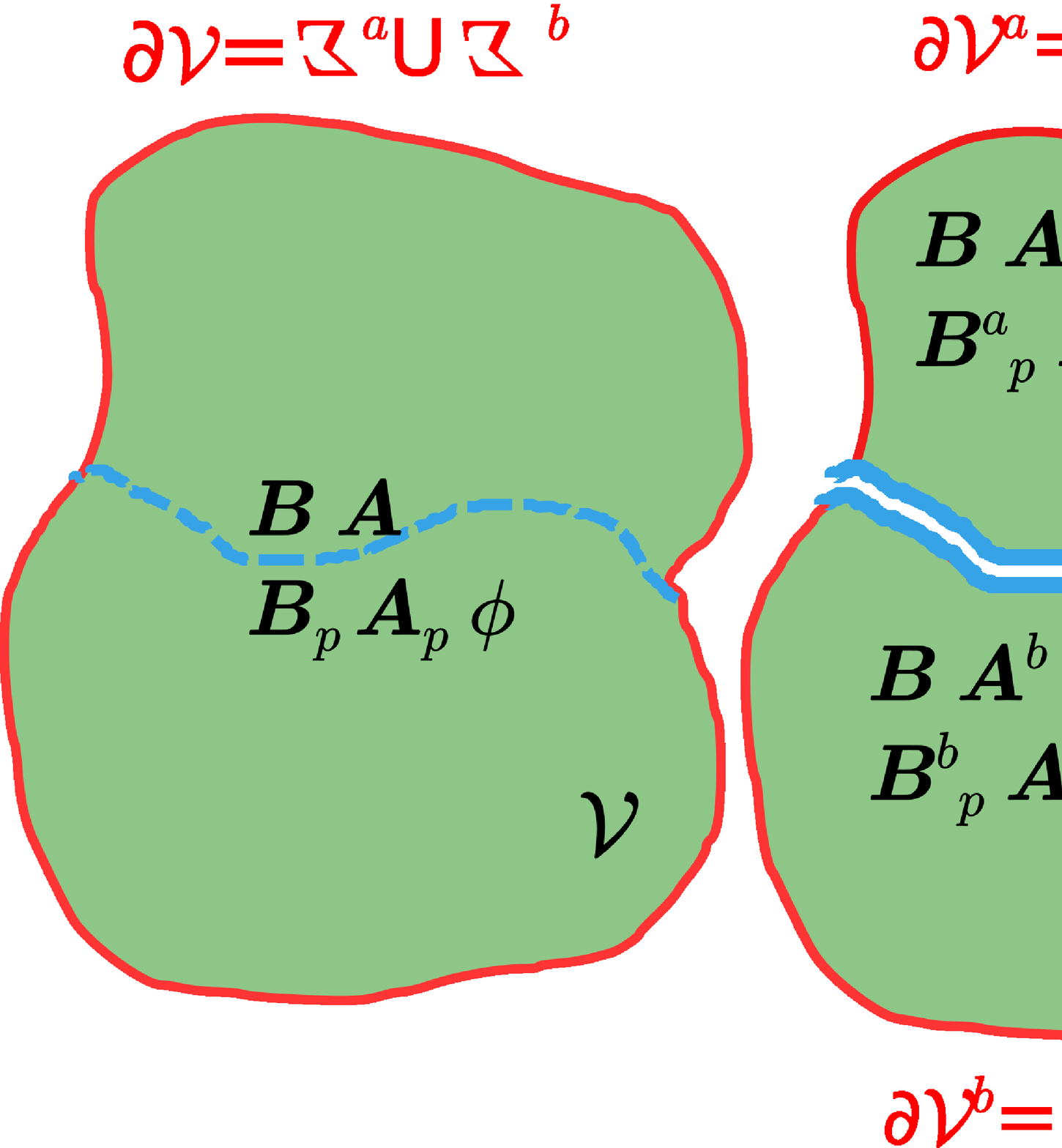}\\   %.eps
  \caption{Sketch of the volume splitting: an interface $\surfS$ splits the finite volume $\vol$ bounded by $\surf$ into two subdomains, $\vola$ and $\volb$, each one bounded by the surface $\surfa$ and $\surfb$, respectively.
   }
  \label{fig:split_v}
 \end{figure}
%----------------------------------------------------------------------------------------------

%%%%%%%%%%%%%%%%%%
\subsection{Volume partition}\label{s:volumes}
We consider the case of two contiguous volumes of finite size and $\vol=\vola \cup \volb$, such that $\vola$ and $\volb$ are bounded by the surfaces $\surfa$ and $\surfb$ with external normals $\hatna$ and $\hatnb$, respectively. 
\Fig{split_v} shows a graphical representation of the volumes involved.
The boundary surface of each sub-volume can be split into an interface ($\surfS$) plus a non-interface ($\surfZ$) contribution as %, \ie
\BA
\surfa&=& \surfS^a \cup \surfZ^a \, , \label{eq:split_a}\\
\surfb&=& \surfS^b \cup \surfZ^b \, , \label{eq:split_b}
\EA
with the boundary $\surf$ of the volume $\vol$ given by
\BE
\surf = \surfZ^a \cup \surfZ^b \, . \label{eq:split_v}
\EE
The interface $\surfS^a$ and  $\surfS^b$ represents the same surface but differ for the orientation of the normal, $\hatn^a=-\hatn^b$. 
When the orientation of the normal is not required we drop the superscript from $\surfS$. 
In order to have a more compact notation, we also introduce 
\BE
     Z^{ab} = \left \{ \begin{array}{l l}
                   Z^a  & \quad \forall\ \vx \in \vola \\
                   Z^b  & \quad \forall\ \vx \in \volb
                 \end{array}  \right . \label{eq:Zab}
\EE
where $Z$ is any function or vector defined separately in $\vola$ and $\volb$. 

All volumes are assumed to be simply connected, to avoid the difficulties of multi-valued gauge functions, but no assumption is made on the shape of the volumes or of the interface. 
No further assumption is made at this point about the geometry of the system.  

For each of the three considered volumes $\vol$, $\vola$ and $\volb$, the relative magnetic helicity, \eq{Hv}, can be computed.
The question that we wish to address in this section is: what is the general relation between the three correspondent relative helicity values $\Hv (\vB,\vol)$, $\Hv (\vB,\vola)$, and $\Hv (\vB,\volb)$?

%%%%%%%%%%
\subsection{Difference in the reference fields} \label{s:pot_example}
The relative helicity \eq{Hv} is gauge-invariant because the reference potential field $\vBp$ for the full volume $\vol$ is defined by \eq{Nabla_phi_bc}, thus satisfying  the gauge-invariance condition \eq{Bp_bc}. 
Similarly, the computation of the relative helicity for the two subvolumes $\vola$ and $\volb$ requires to define reference fields that fulfil the same condition, \eq{Bp_bc}, but in each subvolume separately.  
In other words, the (potential) reference fields $\vBpa$ and $\vBpb$ are uniquely defined by $\Delta\ \phia =0$  in $\vola$ and the boundary condition 
\BE
\left . \hatna \cdot \Nabla\phia \right |_\surfa = \left . \hatna \cdot \vB \right |_\surfa   \, , \label{eq:Nabla_phia_bc} \\
\EE
and $\Delta\ \phib =0$ in $\volb$ and the boundary condition
\BE
\left . \hatnb \cdot \Nabla\phib \right |_\surfb = \left . \hatnb \cdot \vB \right |_\surfb   \, , \label{eq:Nabla_phib_bc}
\EE
respectively.
Notice that, because of the way the volume $\vol$ is split, the boundary conditions for the above Laplacian equations on the non-interface boundaries $\surfZ^{ab}$ of  $\vola$ and $\volb$ are the same as for $\vBp$ (see %\eqs{Nabla_phi_bc}{split_v} and 
\fig{split_v}).

In order to understand the differences between the reference fields, let us first consider a special case:
if $\hatn \cdot \vB = \hatn \cdot \vBp$ on the interface $\surfS$, then the normal components of $\vBpa$ and $\vBpb$ match that of $\vBp$ on the interface $\surfS$ too.  
Then it follows from the uniqueness of the solution to the Laplace problems that, in this special case, it is  $\vBpa=\vBp$ in $\vola$ and $\vBpb=\vBp$ in $\volb$.  
The combined field $\vBpab$ as defined by \eq{Zab} is continuous across $\surfS$, and we have $\vBpab=\vBp$. 

However, for a generic $\vB$ field, $\hatn \cdot \vB$ is different from $\hatn \cdot \vBp$ on the interface $\surfS$.
Therefore, in the general case, the solutions of the Laplacian equations in $\vola$ and $\volb$ provide potential fields $\vBpa$ and $\vBpb$ that are different from $\vBp$ in each of the subvolumes. 
Moreover, the transverse components of $\vBpa$ and $\vBpb$ are in general different on both sides of the interface $\surfS$.   
The corresponding field $\vBpab$ in the full volume $\vol$ is not fully potential but it contains a current sheet on $\surfS$.   
As we show in the next section, this difference between the reference field in $\vol$ and in the subvolumes $\vola$ and $\volb$ is at the core of the non-additivity of relative magnetic helicity.

It is worth noting that the difference between $\vBp$ and $\vBpa$ (respectively, $\vBpb$) in $\vola$ (respectively, $\volb$) is not a consequence of the choice of potential fields as reference fields. 
Indeed, the same difference is to be expected for other non-potential reference fields.
This is because, on the one hand, the gauge-invariance condition \eq{Bp_bc} needs to be imposed on the interface $\surfS$ for the reference fields of $\vola$ and of $\volb$.
On the other hand, $\surfS$ is not a boundary of $\vol$, and therefore the reference field of $\vol$ cannot be specified on $\surfS$.
We conclude that, in general, the reference field $\vBp$ in $\vol$ is not derived from the same information as $\vBpa$ and $\vBpb$, and therefore $\vBp$ is not simply the  juxtaposition  of the reference fields $\vBpa$ in $\vola$ and $\vBpb$ in $\volb$. 
This, and the resulting discontinuity between $\vBpa$ and $\vBpb$ across $\surfS$ discussed above, are direct consequences of the property of \eq{Bp_bc} that reference fields must fulfil in order for the relative magnetic helicity, \eq{Hv}, to be gauge-invariant.  

Before we consider the relative helicity, let us first briefly discuss the consequences of the volume splitting on the magnetic energy 
\BE
\EB(\vB,\vol) = \frac{1}{2\mu_0} \intv  B^2 \, \dV \, ,
\label{eq:EB}
\EE
and the relative (or free) magnetic energy
\BE
 E(\vB,\vol) = \frac{1}{2\mu_0} \intv \left(B^2 - B_{\rm p}^2\right) \, \dV  \, .  
 \label{eq:E}
\EE 
By introducing the volume splitting of \sect{volumes} to the free energy we have
\BE
 E(\vB,\vol) -E(\vB,\vola)-E(\vB,\volb) 
    = \frac{1}{2\mu_0}  \intv \left( (\Bp^{\rm ab})^2 - \Bp^2 \right) \, \dV  \, .  
 \label{eq:Ediff}
\EE 
Since the right-hand side of \eq{Ediff} is in general non-vanishing, then the free energy in $\vol$ is not simply equal to the sum of the free energies in the composing subvolumes $\vola$ and $\volb$, \ie{} the free energy is a non-additive quantity. 
In particular, the difference of reference magnetic fields in $\vola$ and $\volb$ from the one in $\vol$ implies the non-additivity of the relative energy, while the energy $\EB$ is manifestly additive.  
%%%%%%%%%%%%%%%%%%%%%%
\subsection{Relative Magnetic helicity of contiguous volumes: General formulation}\label{s:main}
Without loss of generality, the relative magnetic helicity in $\vol$ can be formally written as 
\BE
\Hv(\vB,\vol)= \Hv (\vB,\vola) + \Hv (\vB,\volb) +\delta\Hv 
\label{eq:H_add_gen}
\EE
with 
\BA
 \delta\Hv &=& \Hv(\vB,\vol)- \Hv (\vB,\vola) -\Hv(\vB,\volb) \nonumber \\
           &=& \delta\HH -\delta\HH_p+\delta\HH_{\rm mix} 
\label{eq:dHv}
\EA
where  we defined
\BA
\dHH    &=&  \HH(\vB     ,\vol)  -\HH(\vB  ,\vola)      -  \HH(\vB  ,\volb)           \label{eq:dH}\\
\dHHp   &=&  \HH(\vBp    ,\vol)  -\HH(\vBpa,\vola)      -  \HH(\vBpb,\volb)           \label{eq:dHp}\\
\dHHmix &=&  \Hmix(\vB,\vBp,\vol)-\Hmix(\vB,\vBpa,\vola)   \nonumber \\
        &\phantom{=}&                                   - \Hmix(\vB,\vBpb,\volb) \, .\label{eq:dHmix}\
\EA
\Eq{H_add_gen} is the result of a simple reorganization that  collects in $\delta\Hv$ all contributions that make the relative magnetic helicity a non-additive quantity  and, by grouping similar terms together, allows for cancellations between them.
In \appx{corrections} we show that, by using \eqs{H}{HmixS} in \eq{dHv}, we obtain 
\BE
 \dHH  =  \int_\surfS \chi \left ( \vB \cdot \dSa \right ) \, , \label{eq:dH_f}
\EE
\BE
 \DHp=\DHp^{Coul} + \DHp^{Surf}
 \label{eq:dHp_f}
\EE
where
\BA
\dHHp^{Coul} &=& \intva \phi^a \left (\Nabla \cdot \vApa \right ) \dV + \intvb \phi^b \left (\Nabla \cdot \vApb \right ) \dV
      \nonumber \\
             &-& \intv \phi \left (\Nabla \cdot \vAp \right ) \dV
      \label{eq:dHp_coul_f} \, , \\
\dHHp^{Surf} &=& \ints  \left ( \phi \vAp -\phi^{ab}\vAp^{ab} \right ) \cdot \dS  \nonumber \\
             &-& \intsS \left ( \phi^a\vApa -\phi^b\vApb \right ) \cdot \dSa
      \label{eq:dHp_surf_f} \, ,
\EA
and
\BA
\dHHmix
        &=& \ints  \left ( \vA^{ab} \times \left ( \vAp - \vAp^{ab}\right) \right) \cdot \dS   \nonumber \\
        &+& \intsS \left [ \vAa\times \left(\vAp-\vApa \right ) - \vAb\times \left(\vAp-\vApb\right ) \right ]\cdot \dSa      \nonumber \\
        &-& \intsS \chi \left ( \vBp\cdot \dSa \right)  \, ,
      \label{eq:dHmix_f} 
\EA
where we use the notation of \eq{Zab} for all fields defined in the subvolumes $\vola$ and $\volb$, and $\chi$ is the gauge function defined by 
\BE
 \Nabla \chi= \left .\left ( \vAb -\vAa \right ) \right |_\surfS \, ,
 \label{eq:chi_sigma}
\EE
with $\chi$ a function of the interface variables only, see \eq{chi_ab_sigma}.
On the interface $\surfS$, the infinitesimal oriented surface was chosen to be that of $\dSa$.
The study of the properties of \eqss{H_add_gen}{dHmix_f} is the main focus of this article. 

The first and most important result is that \eq{H_add_gen} shows in the most general way that the relative magnetic helicity is not an algebraically additive quantity: The relative magnetic helicity in the entire volume $\vol$ is not simply the sum of the relative helicity of the composing subvolumes $\vola$ and $\volb$, but a general non-vanishing additional term, $\delta\Hv$, is present.
Note that, since the left-hand side (LHS) and the first two terms on the right-hand side (RHS) of \eq{H_add_gen} are gauge-invariant, then $\dHv$ must be globally gauge-invariant too. 
\appx{gi-additivity} outlines how to see this directly from the terms in $\delta H$.

The gauge-invariance of \eq{H_add_gen} implies that the non-additivity of relative magnetic helicity is a general property: a special gauge that makes the relative helicity additive (or even partitionable between volumes) does not exist.
This does not rule out that a very special combination of geometry, choice of the reference field, and boundary conditions may exist in which relative magnetic helicity is additive, but this is not true in general.

Finally, we notice that \eqss{dH_f}{dHmix_f} contain three types of terms in the representation that we have chosen, namely volume, interface, and outer boundaries (or non-interface) surface terms. 
This formalism allows for a more direct treatment of specific limits in the next sections, but is by no means the only possible one. 
Let us now  discuss the individual non-additivity terms. 

\subsubsection{$\dHH$ : Non-additivity of the magnetic helicity} \label{s:dH_discuss}
The non-additive term $\dHH$ of \eq{dH_f} is an interface term that, for arbitrary $\vB$, depends solely on the gauge specification, and can be therefore eliminated by specific gauge choices for $\vAa$ and $\vAb$ that insure $\chi=0$ on $\surfS$, see \sect{gauge_practical}.
\subsubsection{$\dHHp$ : Non-additivity of the helicity of the reference fields}\label{s:dHp_discuss} 
The $\dHHp$ in \eq{dHp_f} is composed of a volume and surface terms.
The volume term, $\dHHp^{Coul}$ contains the Coulomb gauge conditions for the three reference fields, a gauge that can be chosen to have this term vanish. 
It is interesting to note that the Coulomb conditions appear explicitly only in relation to reference fields, and not for any of the other vector potentials.
The same happens for the time evolution of the relative magnetic helicity in Eq.~(25) of \cite{Pariat2015a}, and in Eqs.~(13,41) regulating the evolution of the current-carrying and volume-threading relative magnetic helicity derived by \cite{Linan2018}. 
All these cases express helicity contributions due to sources in the vector potentials of the potential fields, and thereby in the helicity, of the reference potential fields. 

While $\dHHp^{Coul}$  accounts for volume differences, the surface term $\dHHp^{Surf}$ in \eq{dHp_surf_f} contains interface and non-interface contributions that depend on the components of vector potentials of the reference fields that are normal to the boundaries, and on the scalar potentials of the same reference fields.
Since the reference fields in $\vol$ and $\vola$ ($\vol$ and $\volb$, respectively) do \textit{not} represent the same field (see \sect{pot_example}), there is no general gauge relation between the vector potentials $\vAp$ and $\vApa$ ($\vAp$ and $\vApb$, respectively) that can be used to simplify these expressions.  

The last term in \eq{dHp_surf_f} is an interface term accounting for the discontinuity of the transverse components in the reference fields at $\surfS$.
As discussed in \sect{pot_example}, this term can be seen as the contribution due to a surface current generated by the discontinuity of the transverse components of $\vBpa$ and $\vBpb$ across $\surfS$, see also \sect{BF_additivity}.

\subsubsection{$\dHHmix$ : Non-additivity of the mixed helicity}\label{s:dHmix_discuss}
The first two integrals in \eq{dHmix_f} are directly related to the transverse components of the vector potentials at the boundaries.
The additional complication of \eq{dHmix_f} with respect to the simpler \eq{HmixS} is that such integrals involve cross interactions between different vector potentials. 

The last term in \eq{dHmix_f}  is similar to that in $\dHH$, but involves $\vBp$ rather than $\vB$, and similar considerations hold.
Since, in general, $\vB$ and $\vBp$ differ on $\surfS$, then the combination of these two terms is nonzero, and it is related to the current-carrying part of the field, $\vBj$. 
Unless the gauge choices for $\vAa$ and $\vAb$ ensure $\chi=0$, the only other case where the two terms cancel each other is that $\vB=\vBp$ on $\surfS$, which, according to the discussion in \sect{pot_example}, is a very special case. 

   % Case of helicity/energy
In summary, the non-additivity of the relative helicity $\Hv$ has the same origin as that of the relative energy $E$, a difference of reference field in each subvolume with the one in the full volume. 
This is the case even when the lowest energy state, the potential field, is selected as reference field.  
Still, the non-additivity terms of $\Hv$ are much more complex than the one for $E$, as they also involve the vector potentials.

%%%%%%%%%%%%%%%%%%%%%%
\section{Applications of the partition equation with specific gauges}\label{s:gauges} 
\Eq{H_add_gen} is a gauge-invariant, general expression of the relative helicity partition that does not make any assumption about the specific gauges and boundary conditions that are used to compute the vector potentials. 
Such specifications are however required for its practical application. 

The constraints on the scalar and vector potentials defined so far derive basically from the gauge-invariance constraint, see \eq{Bp_bc} and \sect{pot_example}. 
In particular, the scalar potentials are determined by solving the Poisson problems \eqs{Bp_def}{Bpab_def} that define $\phi$, $\phia$, and $\phib$ each modulo a constant. 
The vector potentials must obey the curl relations with the fields, \eq{AaAb}.
In addition, the connection between vector potentials of different volumes are prescribed by \eq{chi_sigma} (or, more specifically, \eq{chi_ab_sigma}).
This set of constraints is not sufficient to determine the vector potentials.

A gauge should be properly defined as the set of equations and boundary conditions that uniquely determines the scalar and vector potentials for a given magnetic field $\vB$ in $\vol$.
In this sense, in \sect{test} we refer different sets of different boundary conditions for the vector potentials as different gauges.
In more relaxed sense, we often use the term gauge to mean a group of gauges, %subset of that set, 
as when we speak of the ``Coulomb gauge'' meaning the $\divA=0$ condition only. 
This is rather a family of gauges, to which additional boundary conditions must be added to uniquely determine the vector potentials.
 
Insofar they do not conflict with the other gauge constraints, such additional equations and/or boundary conditions are arbitrary, and, thanks to gauge invariance, they do not affect the outcome of \Eq{H_add_gen}.

\subsection{Coulomb  gauge}\label{s:gauge_coul}

Assuming that all three vector potentials are solenoidal, $\Nabla\cdot\vAp=\Nabla\cdot\vApa=\Nabla\cdot\vApb=0$, then $\dHHp^{Coul}=0$. 
In addition, \eq{Aab} in conjunction with the Coulomb gauge, restricts the possible choice of the gauge functions to the class of functions that satisfy $\Delta \chi^a=\Delta \chi^b=0$.

\subsection{DeVore-Coulomb gauge}\label{s:gauge_devore}

The DeVore gauge \citep{DeVore2000,Valori2012,Moraitis2018} sets one of the components of the vector potential equal to zero. 
For concreteness, let us assume that $\surfS$ is a plane parallel to the $xy$-plane, and set $\Az=0$, as in \cite{Valori2012}. 
The main advantage of the DeVore gauge is that it is very accurate and fast to compute numerically \citep{Valori2016}, since the vector potentials are computed by one-dimensional vertical integration of the magnetic field starting from one of the boundaries. 
A particularly useful formulation of the DeVore gauge requires, in addition, that the two-dimensional integration functions appearing in the computation of the vector potential of the potential field  are  solenoidal (see Section~5 in \cite{Valori2012}). 
In this case, the DeVore gauge ensures that $\Nabla\cdot\vAp=0$, \ie\ it is a DeVore-Coulomb gauge for the vector potential. 
If the DeVore-Coulomb gauge is adopted for all three vector potentials $\vAp$, $\vApa$ and $\vApb$, then also in this case $\dHHp^{Coul}=0$.

\subsection{Boundary conditions}
\label{s:boundary_conditions}
Before analyzing further \eq{H_add_gen}, let us first consider the relative magnetic helicity in a single volume as given in \eq{nongi_decomp}.
A commonly used condition that is often used in combination with the Coulomb gauge for $\vAp$ \citep[see \eg{}][]{Berger1999,Thalmann2011} is that 
\BE
\hatn \times \vA=\hatn\times \vAp \,, 
\label{eq:flux-surface}
\EE
 \ie{} that the vector potential of field and reference field have the same tangential components on the boundary of the considered volume. 
Such a boundary condition is allowed since \eq{flux-surface} implies \eq{Bp_bc}.
In this case, $\Hmix(\vB,\vBp,V)$ vanishes by \eq{HmixS} and the relative magnetic helicity \eq{Hv} equals the difference between the helicity of the field and the helicity of the relative reference field, \ie{}
\BE
\Hv (\vB,\vol) = \HH(\vB,\vol) - \HH(\vBp,\vol)
\label{eq:Hv_noHmix}
\EE
which is the definition of relative helicity predating \eq{Hv}, used \eg{} in \cite{Berger1984} and \cite{Jensen1984}. 
In the practical computation of vector potentials, \eq{flux-surface} is often indirectly imposed by assuming that $\vA=\vAp$ on $\surf$ as a boundary condition for $\vA$. 
We stress, however, that the boundary condition in \eq{flux-surface} is not compatible with the DeVore gauge, as shown by Eq.~(31) of \cite{Valori2012}.
Similarly, it is not possible, in general, to have $\dHHmix=0$ in \eq{dHmix_f} in this gauge.

A special boundary condition is when $\surf$ is a flux surface.
In this case, from $\hatn\cdot\vB=0$, it follows that $\vA$ can be written as $\vA=\hatn A_{\rm{n}} + \Nabla_\perp \chi$ for some function $\chi$, where $\Nabla_\perp$ is the gradient normal to $\hatn$. 
Then, substituting in \eq{HmixS}, we can extend the $\Nabla_\perp\chi$ to a full gradient without changing the integral, and we obtain 
\BE
\Hmix=\int_\surf \Nabla \times (\chi \vAp) \cdot \dS - \int_\surf \chi \vBp\cdot\dS \, ,
\label{eq:Hmix_flux_surface}
\EE
where the last term ion the RHS vanishes since, from \eq{Bp_bc}, $\surf$ is a flux surface of $\vBp$ too.
If the flux surface $\surf$ is  closed, then the first term on the RHS of \eq{Hmix_flux_surface} vanishes too, and $\Hmix$=0 in this case. 
Moreover, since $\hatn\cdot\vB=0$ in \eq{Nabla_phi_bc}, then $\vBp=\vec{0}$, and $\Hv(\vB,\vol)=\HH(\vB,\vol)$. 

If the boundary condition of \eq{flux-surface} is used in the computation of the three pairs vector potentials $(\vA,\vAp)$,  $(\vAa,\vApa)$, and  $(\vAb,\vApb)$, then \eq{dHmix}, with \eq{HmixS} defining $\Hmix$, directly shows that 
\BE
\Hmix(\vB,\vBp,\vol)= \Hmix^a(\vB,\vBpa,\vola) = \Hmix^b(\vB,\vBpb,\volb) =0 \, ,
\EE
and thus $\dHHmix=0$. 
The same result can be also obtained directly from \eq{dHmix_f} using \eq{Aab} to write $\hatn \times (\vApab +\Nabla \chi) =\hatn\times \vAp$.  

\subsection{Computation of the gauge function $\chi$}\label{s:gauge_practical}
The gauge function $\chi$ defined by \eq{chi_sigma} and appearing in \eqs{dH_f}{dHmix_f} can be computed by direct integration, using the fundamental theorem of calculus for line integrals as 
\BE
\chi(\vx)=\chi(\vec{a})+\int_{\mathcal{C}(\vec{a},\vx)} \left (\vAb-\vAa \right ) \cdot \rm{d}\vec{l}  \, ,
\label{eq:chi_integral}
\EE
for any curve $\mathcal{C}(\vec{a},\vx)$ on $\surfS$ connecting points $\vec{a}$ to $\vx$, with $\chi(\vec{a})=0$ as a general prescription.

In general, the transverse components of $\vAa$ and $\vAb$ on $\surfS$ are different, and $\chi$ is a non-vanishing function of the $\surfS$ variables.
However, depending on the gauge, special  boundary conditions on $\vAa$ and $\vAb$ can be imposed such that $\chi=0$.  
We give examples of such boundary conditions in \sect{test} for the DeVore gauge.

%%%%%%%%%%%%%%%%%
\section{Relation with other approaches}\label{s:models}
We show in this section how our general formula \eq{H_add_gen}, in the proper limits, reproduces relevant results on helicity partition known from the literature. 
\subsection{\cite{Berger1984a}'s additivity formula}\label{s:BF_additivity}
In the second part of Section~3 \cite{Berger1984a} derive a relative helicity summation equation for two domains, their Eq.~(45), which in our notation reads
\BA
\HH(\vB,\vola)+\HH(\vB,\volb) &=& \Hv(\vB,\vola)+\Hv(\vB,\volb)\nonumber\\
&+&\HH(\vBpa,\vola)+\HH(\vBpb,\volb).
\label{eq:bfadd}
\EA
This equation is less general than our \eq{H_add_gen} because, first, it assumes that  the combined domain $\vol$ is magnetically closed; second, it adopts the definition of relative magnetic helicity \eq{Hv_noHmix}, rather than the more general \eq{Hv}; third,  \eq{bfadd} is an addition formula, rather than a partition one like our \eq{H_add_gen}, in the sense that \cite{Berger1984a} are not concerned with the general relation to the relative helicity of the total volume (which indeed does not appear in \eq{bfadd}).
In order to relate  \eq{bfadd} to our \eq{H_add_gen} we then assume that (i) $\hatn\cdot\vB=0$ on $\surf$, (ii) both $\vAa\times\hatn=\vAb\times\hatn$ and $\vApa\times\hatn=\vApb\times\hatn$ on the interface $\surfS$, and (iii)  $\vA=\vAa$ in $\vola$ and $\vA=\vAb$ in $\volb$. 
To see that \eq{H_add_gen} reduces to \eq{bfadd} under these conditions, note first that condition (i) implies that $\vBp=\boldsymbol{0}$ and $\Hv(\vB,\vol)=\HH(\vB,\vol)$, see \sect{boundary_conditions}. 
Adding $\Hv(\vB,\vol)$ on both sides of the equation, we can rewrite \eq{bfadd} as
\BA
\dHv = \dHH - \dHHp.
\label{eq:bf_formal}
\EA
This would be equivalent to \eq{H_add_gen} if $\dHHmix=0$. 
To see that this follows from conditions (i) and (ii), first note that $\Hmix(\vB,\vBp,\vol)=0$, see \sect{boundary_conditions}, so that
\BA
\dHHmix &=& - \int_{\partial \vola}\vAa\times\vApa\cdot\dSa- \int_{\partial \volb}\vAb\times\vApb\cdot\dSb,\\
&=& - \int_{\partial \vol}\vAab\times\vApab\cdot\dS,
\EA
where the last step used condition (ii) on $\surfS$. 
From condition (i) we have that $\hatn\times\vAab=\hatn\times\nabla\xi$ for some function $\xi$, so that
\BE
\dHHmix = \int_{\surf}\xi\vB\cdot\dS = 0.
\EE
So Eq. (45) of \cite{Berger1984a} is indeed a special case of our more general \eq{H_add_gen}. 
Finally, we can then formally adopt condition  (iii), which directly results into $\delta \HH=0$ in \eq{bf_formal}.
Under the same conditions (i-iii), \cite{Berger1984a} also observe that the relative helicity becomes an additive quantity if the interface (our $\surfS$) is a planar or spherical surface, since then $\HH(\vBpa,\vola)=\HH(\vBpb,\volb)=0$.
\subsection{The \cite{Longcope2008a} approach}\label{s:Longcope}
The approach in \cite{Longcope2008a} addresses the partition problem explicitly and, in this sense, is the most relevant to compare with. 
The definition of relative magnetic helicity adopted by \cite{Longcope2008a} is that of \eq{Hv_noHmix}, with $\hatn\times\vA=\hatn\times\vAp$ on the boundary of each considered (sub-)volume. 
From  the point of view of the problem formulation, \cite{Longcope2008a} explicitly focus on macroscopic coronal flux tubes  that have their bases in the photospheric plane (footprints) and are laterally bounded in the corona by flux surfaces. 
In our notation, their coronal flux surfaces are our separation interfaces between subvolumes, $\surfS$, whereas their photospheric footprints belong to the $\surfZ$ portion of the boundary of each subvolume. 

For each subvolume, in one of the analysed cases, \cite{Longcope2008a} consider that the reference field  is potential and restricted to the subvolume, yielding the additive self-helicity formula (their Eq.~(16)).
This case shares the same approach (and limitations) to the reference potentials as ours, see discussion in \sect{pot_example}. 
In order to link Eq.~(16) in \cite{Longcope2008a} to our \eqss{dH_f}{dHmix_f}, let us restrict their notation to two subvolumes only.
The first term on the LHS of Eq.~(16) in \cite{Longcope2008a} is then the relative magnetic helicity of the entire coronal volume, \ie{} the LHS of our \eq{H_add_gen}.
The second term on LHS of Eq.~(16) in \cite{Longcope2008a} is the sum of the relative magnetic helicity of the two composing subvolumes, \ie{} it is equal to the first two terms in the RHS of our \eq{H_add_gen}.
Then, we are left to show under which conditions $\dHHmix$ in \eq{dHmix_f} is equal to the RHS of  Eq.~(16) in \cite{Longcope2008a}.
First, let us keep $\dHHp$ in the form given in \eq{dHp}.
Second, adopting the assumption that $\surfS$ is a flux surface in \eqs{dH_f}{dHmix_f} directly yields $\dHH=0$ and 
\BA
\dHHmix    &=& \intsa  \left ( \vAa \times \left ( \vAp - \vApa \right) \right) \cdot \dSa   + \int_{a\rightarrow b} \dots \,,\nonumber \\
           &=& \intsa  \vAp \cdot\left ( \hatna \times \vAa \right ) \ \ds  \nonumber \\ 
           &-& \intsa  \vApa\cdot\left ( \hatna \times \vAa  \right ) \ \ds                     +\int_{a\rightarrow b} \dots \,,
\label{eq:lm_1}
\EA
where the last integral indicates the repetition of all integrals with index $a\rightarrow b$.
Since $\surfS$ is a flux surface, $\dHHmix$ vanishes there, see \sect{gauge_coul}.
If we then recall that $\hatna \times \vAa = \hatna \times \vApa$ on the photospheric $\surfZ$ boundary, regrouping $\dHHmix$ and $\dHHp$ terms,  we have
\BA
\dHH &=& \dHHmix  + \dHHp \nonumber \\
     &=& \intva \left(\vApa\cdot\vBpa - \vAp\cdot\vBp \right ) \ \dV + \intsa  \left ( \vApa \times \vAp\right) \cdot \dSa    \nonumber \\
     &+& \int_{a\rightarrow b} \dots \,,
\label{eq:lm_2}
\EA 
which is the RHS of  Eq.~(16) in \cite{Longcope2008a} written in our notation.
Therefore, for the special case of two single subvolumes bounded in the corona by flux surfaces, \eqss{dH_f}{dHmix_f} reduce to the self-helicity expression of Eq.~(16) in \cite{Longcope2008a}. 
%%%%%%%%%%%%%%%%%%%%%%%%
\section{Numerical verification of the partition equation}\label{s:test}
%---------------------------------------------------------------------------------------------
 \begin{figure}[t]
  \centering
  \includegraphics[width=\columnwidth]{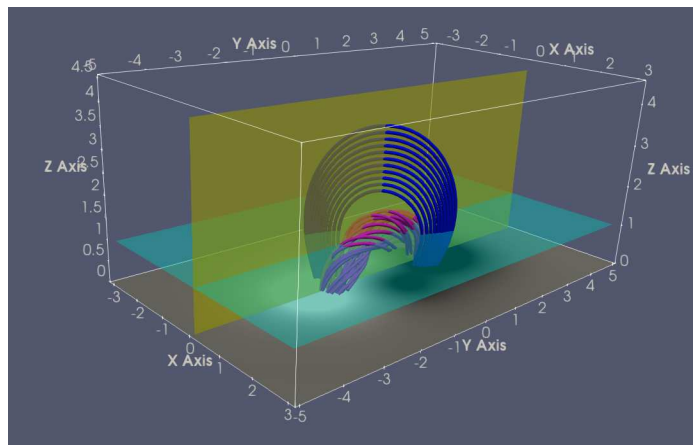}\\   %.eps
  \caption{Selected field lines of the TD equilibrium depicting the flux rope (pink) and the surrounding potential field (blue). 
          The two section planes $\surfS$ used in \tab{gauge} are the $z=1$ plane (cyan) and the $x=0$ plane (yellow), see \sect{test} for details. 
          The distribution of the vertical field component at $z=0$ is shown in greyscale at the bottom.}
  \label{fig:fl_td}
 \end{figure}
%---------------------------
 \begin{table*}
  \caption{Numerical verification of  the partition equation}
  \label{t:gauge}
  \centering
  \small
  \strtable
  \begin{tabular}{@{~}c  c@{\quad}  c@{\quad}  c@{\quad}  c@{\quad}  c@{\quad} c@{\quad} c@{\quad} c@{\quad} c@{\quad} c@{\quad} c@{\quad}}
  \hline
%  case  plane            H_Va             H_Vb                dHr            H_Vab              H_V           Delta_H_V%    bot_Ap   bot_A   dVC_Ap    dVC_A 
  Test &  $\surfS$  &  $\Hv(\vB,\vola)$ & $\Hv(\vB,\volb)$ & $\delta\Hv$ & $H_{\rm sum}$ & $\Hv(\vB,\vol)$ & $\epsilon$ & \bcAp{} & \bcA{} & \dVCAp{} & \dVCA{} \\
  \hline
     1 &   z=1  &      6.4022  &  0.3675  &  0.3887  &  7.1584   &  7.2069  & 0.67    &     t t b   &   t t b   &   y n y   &   n y n  \\
     2 &   z=1  &      6.4101  &  0.3767  &  0.3756  &  7.1625   &  7.2052  & 0.59    &     t t b   &   t t b   &   y y y   &   y y y  \\
     3 &   z=1  &      6.3943  &  0.3582  &  0.3988  &  7.1514   &  7.2087  & 0.79    &     t t b   &   t t b   &   n n n   &   n n n  \\
     4 &   z=1  &      6.4101  &  0.3771  &  0.3764  &  7.1637   &  7.2052  & 0.58    &     t t t   &   t t b   &   y y y   &   y y y  \\
     5 &   z=1  &      6.4045  &  0.3672  &  0.3476  &  7.1193   &  7.1730  & 0.75    &     b b t   &   t t b   &   n n n   &   n n n  \\
     6 &   z=1  &      6.4218  &  0.3771  &  0.3443  &  7.1433   &  7.1840  & 0.57    &     b b t   &   t t b   &   y y y   &   y y y  \\
     7 &   z=1  &      6.3943  &  0.3582  &  0.3988  &  7.1514   &  7.2087  & 0.79    &     t t b   &   t t b   &   n n n   &   n n n  \\
     8 &   z=1  &      6.4220  &  0.3767  &  0.2638  &  7.0625   &  7.2052  & 1.98    &     t t b   &   t b b   &   y y y   &   y y y  \\
     9 &   z=1  &      6.4047  &  0.3582  &  0.2852  &  7.0481   &  7.2087  & 2.23    &     t t b   &   t b b   &   n n n   &   n n n  \\
      \hline                                                                 
   10  &   x=0  &      0.9215  &  1.0414  &  5.2347  &  7.1976   &  7.2052  & 0.11    &     t t b   &   t t b   &   y y y   &   y y y  \\
   11  &   x=0  &      0.9215  &  0.9638  &  5.3095  &  7.1948   &  7.2052  & 0.14    &     t t t   &   t t b   &   y y y   &   y y y  \\
   12  &   x=0  &      0.9215  &  1.0414  &  5.2347  &  7.1976   &  7.2052  & 0.11    &     t t b   &   t t b   &   y y y   &   y y y  \\
   13  &   x=0  &      0.9662  &  1.0414  &  5.1834  &  7.1910   &  7.2052  & 0.20    &     t t b   &   t b b   &   y y y   &   y y y  \\
   14  &   x=0  &      0.9720  &  0.7803  &  5.3074  &  7.0596   &  7.2087  & 2.07    &     t t b   &   t b b   &   n n n   &   n n n  \\
  \hline
  \end{tabular}\\
\tablefoot{Numerical verification of \eq{H_add_gen} using the TD test sliced with a plane $\surfS$.
           The column Test labels the different test cases; 
           $\surfS$ is the plane interface separating $\vola$ and $\volb$; 
           $\Hv(\vB,\vola)$ and  $\Hv(\vB,\volb)$ are the relative magnetic helicities of the subvolumes $\vola$ and $\volb$, respectively, whereas $\delta\Hv$ is the non-addictive term; 
           $H_{\rm sum}$ and $\Hv(\vB,\vol)$ are the RHS and LHS of \eq{H_add_gen}, respectively; 
           $\epsilon$ is the error in percentage between  $\Hv(\vB,\vol)$ and $H_{\rm sum}$ as defined in \eq{rel_error};  
           \bcAp{} (respectively \bcA{}) is a triplet representing the integration direction for the computation of the vector potential $\vAp$ (respectively $\vA$) for the volume $\vol$,$\vola$, and $\volb$, respectively; 
           similarly, \dVCAp{} (respectively \dVCA{}) is a triplet representing the boundary condition for the computation of the vector potential $\vAp$ (respectively $\vA$)  on  \bcAp{} (respectively \bcA{}), for the volume $\vol$,$\vola$, and $\volb$, respectively. 
           See \sect{test} for additional details.
}
\end{table*}

In this section the partition formula  \eq{H_add_gen} is verified numerically using the Titov and D\'emoulin model of a bipolar active region \citep[][hereafter TD]{Titov1999}.
\subsection{Numerical model}
\label{test_Model}
%----------------------------------------------------------------------------------------------
The TD model is a parametric solution of the force-free equations that consists of a portion of a circular twisted flux rope embedded in a potential field. 
The specifications of the considered Cartesian volume and the parameters of the particular solution employed here are the same as the N=1 case in Table~3 of \cite{Valori2016}, except for the opposite sign of the twist. 
\Fig{fl_td} shows selected field lines depicting the flux rope and the two sectioning planes discussed below.

The computation of the vector potentials is performed here using the DeVore gauge $\Az=0$, see \sect{gauge_devore}, as implemented in \cite{Valori2012}.
The method has two parameters, representing different gauges of the DeVore family, as follows. 
First, a one-dimensional integral in the z-direction is involved in the computation of the vector potentials.  
This integral can be performed starting from either the bottom (bc=b) or from the top (bc=t) of the considered volume, corresponding to Eq.~(10) and Eq.~(11) of \cite{Valori2012}, respectively. 
Second, two different boundary conditions can be used in the computation of the vector potential at the starting boundary, namely Eqs.~(24,25) or Eq.~(41) in \cite{Valori2012}. 
As discussed in \sect{gauge_devore}, the latter applied to a potential field results in the DeVore-Coulomb (dVC) gauge.
We then use the notation dVC=n (no) and dVC=y (yes) if, respectively, Eqs.~(24,25) or Eq.~(41) in \cite{Valori2012} are used. 
Therefore, for each vector potential and volume, a different combination of bc and dVC can be used, effectively testing the gauge dependence of the computed quantities.
There are 4 possibilities for each of the six vector potentials, yielding $4^6=4096$ possible combinations.

In the following we provide few representative examples of the possible gauge combinations for each realization of volume splitting.
For instance, for the test number 2 in  \tab{gauge}, \bcAp=[t,t,b] (respectively, \bcA=[t,t,b]) means that the computation of $\vAp$ and $\vApa$ (respectively, $\vA$ and $\vAa$) was performed starting from the top boundary in the volume $\vol$ and $\vola$, and from the bottom boundary for $\vApb$ (respectively, $\vAb$) in the volume $\volb$. 
The triplets \dVCAp=\dVCA=[y,y,y] mean that Eq.(41) of \cite{Valori2012} was used for the computation of all vector potentials in the three volumes $\vol$, $\vola$, and $\volb$.

\subsection{Numerical verification of \eq{H_add_gen}}
\label{test_Numerical}
%----------------------------------------------------------------------------------------------

\tab{gauge} summarizes the results of testing \eq{H_add_gen} in two representative realizations of volume splitting: in the first one ($z=1$, in cyan in \fig{fl_td}) the interface is a horizontal plane cutting through the flux rope at approximately the location of the apex of the flux rope axis. 
In this realization, most of the flux rope is contained in the lower subvolume $\vola$, whereas $\volb$ mostly contains potential field.
The second realization ($x=0$, in yellow in \fig{fl_td}) is a vertical plane cutting through the flux rope and approximately containing the flux rope axis.
In this realization, the flux rope is approximately split symmetrically between the two subvolumes.

In \tab{gauge},  $\Hv(\vB,\vol)$  and 
\BE
H_{\rm sum}\equiv\Hv(\vB,\vola)+\Hv(\vB,\volb)+\delta\Hv 
\label{eq:Hsum}
\EE
are, respectively, the LHS and RHS of \eq{H_add_gen}, computed independently, and
\BE
 \epsilon=100*\left(\Hv(\vB,\vol)-H_{\rm sum}\right)/\Hv(\vB,\vol)
 \label{eq:rel_error}
\EE
represents the error of the helicity partition formula \eq{H_add_gen} in percentage. 

In most of the cases in \tab{gauge}, the error $\epsilon$ in the partition formula is less then 1\%, which clearly verifies that \eq{H_add_gen} is correct, and that its numerical implementation is extremely accurate. 
The first three tests in the $z=1$ case (test=1,2,3 in \tab{gauge}) show that the error  $\epsilon$  does not depend on the values of the dVC triplets, \ie{} is similar for both the deVore and the deVore-Coulomb gauges.  
This remains true for different combinations of the boundary condition bc for the six vector potentials (see tests 4 to 7 in \tab{gauge}).
In particular, there is no dependence on the bc value for the vector potential of the potential field.
The exception is for \bcA=b in $\vola$ (see tests 8 and 9), where the error $\epsilon$  is around 2\%. 
This gauge corresponds to an upward integration in the computation of the vector potential $\vAa$, \ie{} to use  Eq.~(10) of \cite{Valori2012} for $\vAa$.
In this case, numerical errors that accumulate in the vertical integration end up affecting the accuracy of the gauge function $\chi$, due to \eq{chi_integral}. 
Such analysis is confirmed by the $x=0$ cases in \tab{gauge} where some of the tests are repeated for the vertical slice of the volume.
Therefore, \tab{gauge} shows that our implementation of \eq{H_add_gen} can account for the helicity partition with an error typically smaller than 1\%. 

In order to attain such an accuracy, the computation of the gauge function $\chi$, \eq{chi_integral}, was found to be particularly sensitive.
For that we computed the line integral analogously to the Appendix~3 of \cite{Valori2013}, \ie\ such that the integral is the numerical inverse operation of the derivation operator (in our case, a second order central difference scheme). 
If, for instance, a trapezoidal scheme is used instead, then $\epsilon$  can be easily one order of magnitude larger, or even two in some cases.

In terms of relative magnitude, the values of $\delta\Hv$ compared to the total helicity $\Hv(\vB,\vol)$, \ie\ with respect the LHS of \eq{H_add_gen}, are only about 6\% in the $z=1$ case, but as large as 74\% in the $x=0$ case.
This is a first evidence that the relative importance of the non-additive term for a given field depends on how the volume is sliced, and that it can be very significant indeed.

Within a given case, each line in \tab{gauge} corresponds to a different combination of bc and dVC for the six vector potentials, \ie{} to a different gauge. 
Hence, an estimation of the error for \eq{H_add_gen} in fulfilling  gauge-invariance can be obtained as the standard error of the mean of $\delta\Hv$ values.
Such a statistical error, relative to the mean of $\delta\Hv$,  is 5\% in the $z=1$ case, where $\delta\Hv$ values are relatively small, and 0.5\% for the $x=0$ case. 
Such small variations confirm numerically the gauge-invariance of \eq{H_add_gen}, within the gauge-invariant accuracy of the underlining helicity computation method (see also the accuracy tests in \cite{Valori2016}).  

As anticipated in \sect{gauge_devore}, some of the gauge combinations in \tab{gauge} would allow for analytical cancellations in \eqss{dH_f}{dHmix_f}. 
For instance, the second and third tests correspond to the condition $\vApa=\vAa=\vApb=\vAb$ on $\surfS$ for identical \dVCAp{} and \dVCA{} triplets, which would cancel the first interface term in $\dHHmix$, \eq{dHmix_f}, and set $\chi=0$ from \eq{chi_integral}, yielding $\dHH=0$ in \eq{dH_f} and cancelling the last term in \eq{dHmix_f}.  
Similarly, each time a `y' is present in the \dVCAp{} triplet in \tab{gauge}, then the DeVore-Coulomb gauge is imposed on one of the vector potentials, and the corresponding term in $\dHHp$ should vanish. 
We verified that this is indeed the case to high numerical precision, but the numbers in \tab{gauge} are computed always including all terms in \eqss{dH_f}{dHmix_f}.
Therefore, they account also for small numerical errors deriving, for example, from a non-perfect solenoidal property of the vector potentials.
%----
\subsection{Dependence on the interface position}
\label{test_application}
%----------------------------------------------------------------------------------------------
 \begin{figure}[t!]
  \centering
  \includegraphics[width=\columnwidth]{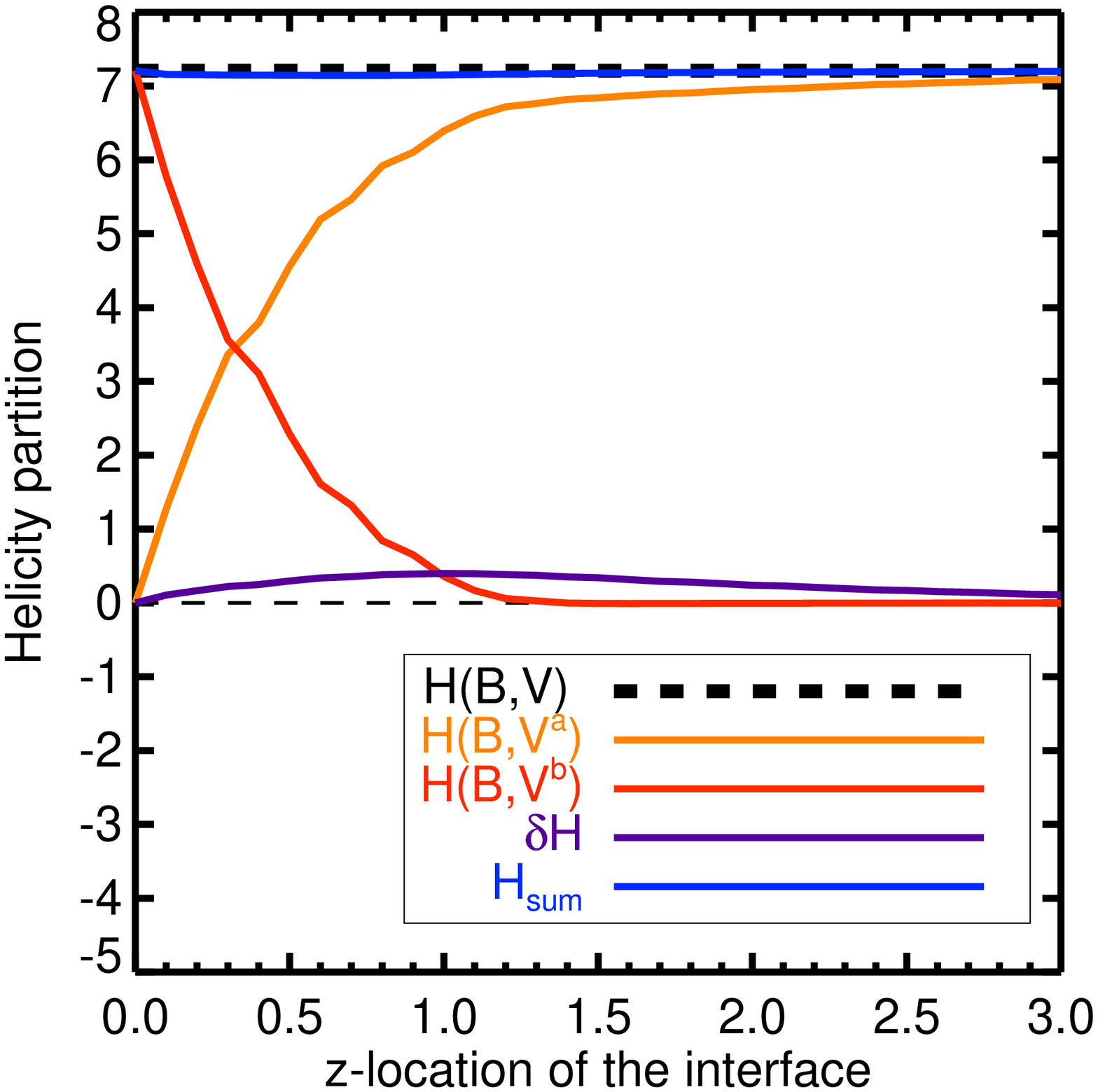}\\   %.eps
  \includegraphics[width=\columnwidth]{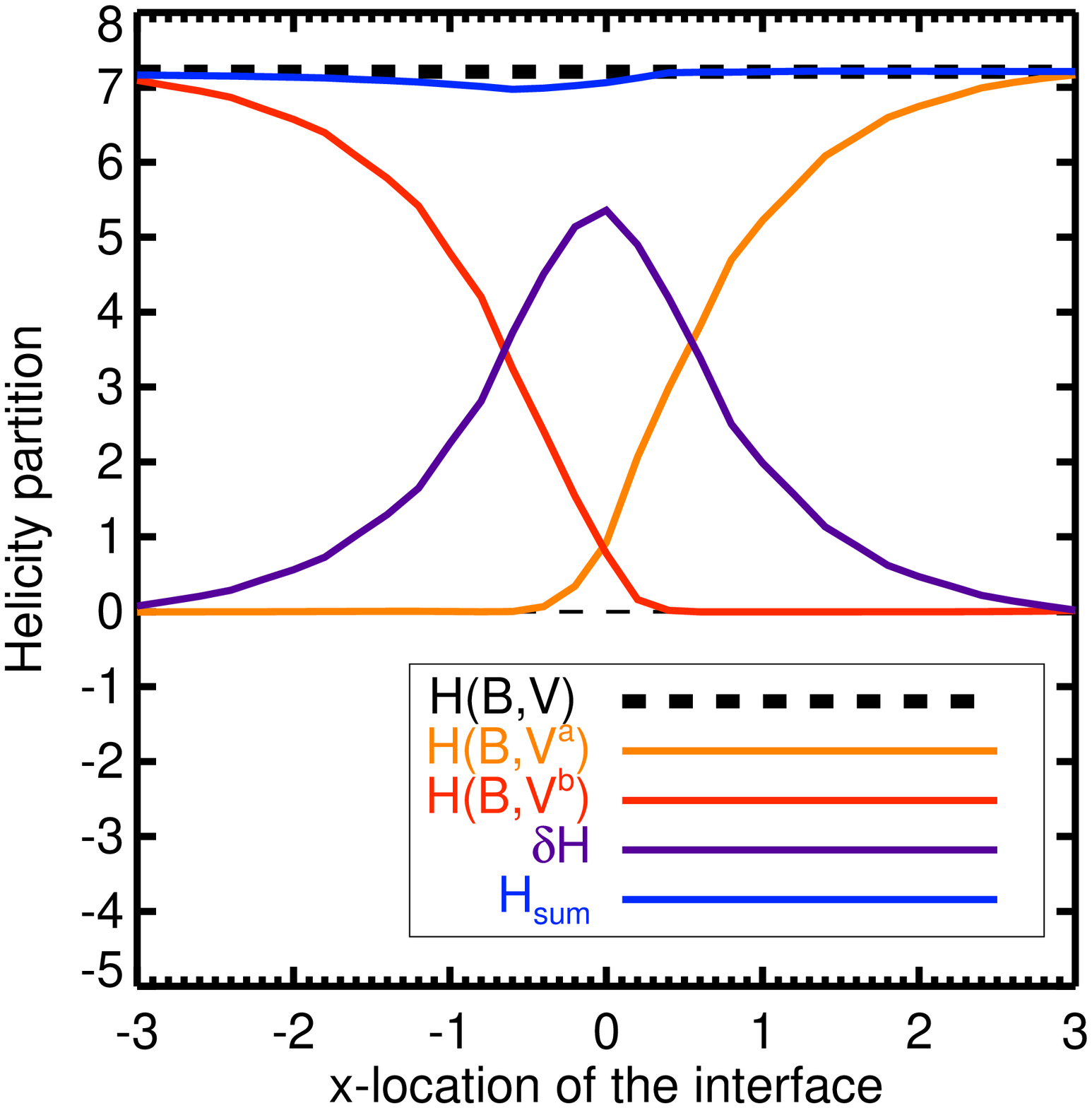}
  \caption{Helicity partition, \eq{H_add_gen}, for the TD volume split with an interface plane perpendicular to the $z$-axis (top panel) and to the $x$-axis (bottom panel), as a function of the interface position. 
   Symbols are the same as in \tab{gauge}, see also \sect{test} for details.}
  \label{fig:td_h}
 \end{figure}
 %----------------------------------------------------------------------------------------------
%
As a first application of the partition formula, \fig{td_h} shows the dependence of the different terms in \eq{H_add_gen} as a function of the position of the interface.
With reference to \tab{gauge}, the gauge used for this application is the same as in test number 3, \ie{} \bcAp=\bcA=[t,t,b] and \dVCAp=\dVCA=[n,n,n].

The top panel of \fig{td_h} refers to a case where the interface is a plane perpendicular to the $z$-axis.
As the interface height changes from z=0 to z=3, its intersection with the TD flux rope rises from the legs, through the apex of the flux rope's axis at $z=1$, until the interface is above the flux rope's top.  
To some extent, this numerical experiment is relevant to the study of flux emergence, as it simulates, for decreasing $z$, the idealized kinematic emergence of a twisted flux tube into the coronal volume, $\volb$.

As the height of the interface raises (top panel of \fig{td_h}), the helicity $\Hv(\vB,\vola)$ of the lower volume $\vola$ (orange curve) and that $\Hv(\vB,\volb)$ of the upper volume $\volb$ (red curve) evolve almost perfectly anti-symmetrically: as $\vola$ includes more and more of the flux rope, its helicity $\Hv(\vB,\vola)$ increases, whereas $\Hv(\vB,\volb)$ decreases of a comparable amount. 
When the interface plane is placed at $z=0.3$, the helicity is  almost equally distributed between the two  subvolumes.
The non-additive term $\delta\Hv$ (violet curve) is always small for all heights of the interface, with a maximum of 6\% of $\Hv(\vB,\vol)$ at $z\simeq 1$.
The accuracy of \eq{H_add_gen} is basically unaffected by the interface position, and $H_{\rm sum}$ (blue curve) overlaps $\Hv(\vB,\vol)$ (dashed black curve) for all positions of the interface.
On the grounds of this first experiment, one would be tempted to say that the non-additivity term $\delta\Hv$ tends to be significantly smaller than the helicity of the composing subvolumes. 

The bottom panel of \fig{td_h} shows a similar experiment  but with a vertical plane (perpendicular to the $x$-axis) that shifts from one side to the other of the flux rope. 
The change in the helicity of the subvolumes in this case is very different, as it involves a significant variation of $\delta\Hv$ too.
As the interface position moves from $x=-3$ to $x=0$, $\vola$ increases at the expense of $\volb$.
In this interval, $\Hv(\vB,\volb)$ (red curve) decreases of the same amount that $\delta\Hv$ (violet curve) increases, whereas $\Hv(\vB,\vola)$ (orange curve), which contains only potential field, is zero, until $x\simeq -0.3$ where it starts rapidly rising. 
The TD solution is line-symmetric with respect to the $z$-axis, therefore a symmetric evolution is present for $x>0$ in the bottom panel of \fig{td_h}. 
Indeed, as the interface moves through $x=0$ towards $x=1$,  $\Hv(\vB,\vola)$ contains more and more of the flux rope and its helicity  $\Hv(\vB,\vola)$ increases, mostly at the expense of $\delta \Hv$. 
Symmetrically to the left part of the plot, as soon as the interface moves out of the flux rope, approximately at $x=0.3$, $\Hv(\vB,\volb)$ is practically zero.

Contrary to the horizontal slicing case in the top panel of \fig{td_h}, in the vertical slicing case in the bottom panel, $\delta\Hv$ is of the same order as $\Hv(\vB,\vol)$ for a large interval of the slicing position, and even several times larger than both $\Hv(\vB,\vola)$ and $\Hv(\vB,\volb)$ in the central interval: in this case, the non-additivity term $\delta\Hv$ is basically never negligible.     
Therefore, depending on the way a volume is sliced, the relative importance of the non-additive term  $\delta\Hv$ can vary significantly, and cannot in general be neglected.

From the discussion in \sect{pot_example} we know that the non-additivity is related to the difference between the reference fields in the subvolumes $\vola$ and $\volb$ with respect to the reference field in the full volume $\vol$.
On the other hand, the position and orientation of the interface directly determines the boundary conditions for the reference fields.
Therefore, the magnitude of $\delta \Hv$ in the two cases in \fig{td_h} is possibly determined by the way in which the boundary conditions for the subvolumes' reference fields change, as a function of the interface position.  
However, to validate such a speculation requires to study the different non-gauge-invariant contributions to $\delta \Hv$ in \eqss{dH_f}{dHmix_f} as a function of the interface orientation and position, a task that we reserve for future studies, see \sect{applications}.

\section{Conclusions}\label{s:conclusions}
%%%%%%%%%%%%%%%%%%%%%%%
\subsection{Results}
The purpose of this work is to study the non-additivity of the relative magnetic helicity in finite-volumes, here formulated as a partition problem between contiguous subvolumes.
In particular: 
\BI
 \item We derive in \sect{main} the general equation for the partition of relative magnetic helicity in a finite-size volume between two contiguous subvolumes separated by an interface, \eqss{dH_f}{dHmix_f}. 
       The explicit assumption of finiteness of the considered volume makes the partition equations directly applicable to numerical simulations. 
       No assumption is made on the shape of interface or the (sub)volumes, as long as they are simply-connected.
       Therefore, \eqss{dH_f}{dHmix_f} can be easily adapted to different geometries, like spherical wedges and the fully spherical case, taking the due care about the required periodicity of vector  potentials (\ie{} barring any mean field in the periodic direction \citep{Berger2003}). 
 \item We show in the  most general way that relative magnetic helicity is not an algebraically additive quantity, and that the non-additive term is gauge-invariant. 
       This allows us to link the non-additivity to the very definition of helicity as relative to a reference field, see \sect{discussion}. 
\EI
 
 Next, we further apply our general equations to specific gauges, used in previous studies and numerical computations, as follows.      
\BI
 \item In \sect{gauges} we analyze the adaptation of the general partition equations to commonly used gauges (Coulomb and DeVore-Coulomb) and boundary conditions often used in the computation of the vector potentials.
 \item In \sect{models} we relate our general approach to well-known reference approaches in the literature, such as those by \cite{Berger1984a} and \cite{Longcope2008a}, which  are obtained under more restrictive assumptions on volumes and gauges. 
In particular, our approach generalizes the one in \cite{Longcope2008a} in few aspects, since all assumptions about the adopted gauge and boundary conditions are relaxed here.
In the first place, we use the definition of relative helicity, \eq{Hv}, rather than a simple difference of helicities, \eq{Hv_noHmix}. 
Second, the assumption that subvolumes are bounded by coronal flux surfaces made in  \cite{Longcope2008a} implies that the volume must be partitioned in a way that the flux is balanced within the photospheric footprint of each subvolume separately.
While this might be a natural way of splitting a coronal volume into a collection of photospherically-anchored flux tubes, this might be  not an easy task in other types of simulation where the logical split of volumes would not necessarily be following flux surfaces.
 \item Finally, we implement and test the accuracy and gauge-invariance of the partition equation using the family of DeVore gauges in \sect{test}, applied to the \cite{Titov1999} solution of the nonlinear force-free equations. 
       These preliminary tests, besides their verification purpose, show that the non-additive term is in general non-negligible, and that it can be significantly larger than the relative helicity of the composing subvolumes in some cases.
       However, these tests also show that the magnitude of the non-additive term depends on the way the volume is split. 
       Therefore, applications of our general formalisms can be devised to investigate under which specific conditions the relative magnetic helicity may become approximately additive (see \sect{applications}). 
\EI

\subsection{Discussion}\label{s:discussion}
The fundamental reason of the non-additivity of relative magnetic helicity lies in its very definition as relative to a reference field.
The very same condition that is needed to ensure the gauge-invariance of relative magnetic helicity, \eq{Bp_bc}, is also responsible for the interface discontinuities in the reference fields that ultimately cause the non-additivity.
This is even more evident when the effect of the finiteness of the considered volume must be considered, as in numerical simulations \citep[see, \eg ][]{Valori2012}.
We stress that the non-additive term $\dHv$ cannot be interpreted as simply the mutual or linking term between the subvolumes, for the same reason that the $\Hmix$ term in the definition of the relative magnetic helicity \eq{Hv} is not the linking between the input and reference field.

On the other hand, as mentioned in \sect{intro}, there are several examples of astrophysical plasmas  where the conservation of magnetic helicity is expected to be a key for understanding the complex processes at a fundamental level, such as the relation between solar and stellar dynamos and the  emergence of magnetic flux through the photosphere, the stability of coronal structure, or the relation between solar eruptions and interplanetary coronal mass ejections. 
In most of such cases, either because of instrumental or numerical limitations, the helicity budget involves volumes that are neither unbounded, nor bounded by flux surfaces, and a general, finite-volume approach is unavoidable.  
The non-additivity of relative magnetic helicity in finite-volumes that we analyse in this work poses a serious threat to the applicability of the conservation of relative magnetic helicity in such fundamental applications. 
As a minimum, our work shows that, when considering the partition of relative magnetic helicity in such applications, the relative magnitude of its non-additive part must be considered.

On a general level, therefore, one would desire to have a different definition of relative magnetic helicity that has additivity \textit{and} gauge-invariance as a core requirements, which, according to the results in this work, is not possible in general.
This impediment does not depend on the \textit{type} of reference field: as shown in \sect{pot_example}, any reference field would lead to the same non-additivity problem because of the boundary conditions that need to be imposed on the interface in order to ensure gauge-invariance.  

It is worth noting that the additivity problem can be solved if one is prepared to relax the requirement of gauge invariance and explicitly fix a gauge in the definition of helicity, since this dispenses with the need for a reference field. 
The original magnetic helicity $\HH(\vB,\vol)$ is then (trivially) additive between sub-volumes, whatever the gauge of $\vA$. 
However, for this additivity to be useful, one ought to be able to compute the helicity of each subregion locally, whereas in general $\vA$ must be computed globally (as, \eg{} with the Coulomb gauge). 
In fact, this problem can be avoided if the interfaces between subdomains are planes or spherical surfaces, since then one can determine a vector potential from $\vB$ purely by integration within these surfaces. 
This is the approach of both \cite{Prior2014a} and \cite{Berger2018}, who describe particular vector potentials for such configurations. 
In \cite{Prior2014a}, the interfaces are parallel planes, whereas  \cite{Berger2018} allow for any spherical nested surfaces. 
In both cases, the corresponding $\HH(\vB,\vol)$ is additive between subvolumes and locally computable. 
Moreover, these authors show that their gauge choices give a particular physical interpretation to $\HH(\vB,\vol)$, which is lacking for an arbitrary choice of gauge.

\subsection{Future applications}\label{s:applications}
Case studies may reveal that the non-additive term is, in fact, almost negligible in specific conditions.
An example of such a case is the kinematic emergence of a flux rope of \sect{test}, and top panel of \Fig{td_h}. 
The bottom panel of the same figure, however, shows that this is not true in general.
The reason for such a difference in the magnitude of the non-additive term is worth being further investigated. 
In other words, there might be specific arrangements of fields and interfaces for which $\delta \Hv$ is indeed non-zero, but still small enough to result into a  relative helicity that is approximately additive. 

A straightforward application of \eqss{dH_f}{dHmix_f} is to characterise the time evolution of the partition of helicity between sub- and super-photospheric volumes in flux emergence simulations such as, \eg\ \cite{Leake2013}.
Such a study can be extended to include dynamo simulations \citep[see, \eg ][]{Brun2017} and the relation to photospheric fluxes \citep[see \eg ][]{Brandenburg2017}.   
Similarly, the partition between the helicity carried by an ejective instability and that remaining confined at lower altitudes during solar eruptions can be studied using simulations such as \cite{Leake2014a,Pariat2015,Toeroek2018}.

On a more theoretical level, our formalism can be used to study the relation between fluxes at the interface of the partitioned volumes and their relation with helicity conservation \citep[see, \eg ][]{Pariat2015a}. 
Similarly, the relative magnetic helicity proxy recently introduced by \cite{Pariat2017b} was found to be a good marker of eruptivity potential in both numerical simulations \citep{Zuccarello2018} and observed active regions \citep{Moraitis2019,Thalmann2019a}. 
The eruptivity proxy is expressed in terms of the helicity of the current-carrying part of the field.
Intriguingly, the same field appears when combining the \eq{dH_f} and the last term of  \eq{dHmix_f}. 

These are only a few examples of applications of our general approach to the partition of relative magnetic helicity. 
Such applications will help us understanding how the conservation of relative magnetic helicity can practically be used in the interpretation of the evolution of complex physical processes in magneto-hydrodynamics.  

%%%%%%%%%%%%%%%%%%%%%%%%%%%%%%%%%%%%%%%%%%%%%%%%%%%%%%%%%%%%%
\begin{acknowledgements}
GV acknowledges the support of the Leverhulme Trust Research Project Grant 2014-051, the support from the European Union’s
Horizon 2020 research and innovation programme under grant agreement No 824135, and of the STFC grant number ST/T000317/. 
LL, EP, KM acknowledge support of the French Agence Nationale pour la Recherche through the HELISOL project ANR-15- CE31-0001. 
LL, EP and PD acknowledge the support of the french Programme National Soleil-Terre. 
ARY acknowledges financial support from Leverhulme Trust project grant 2017-169 and STFC grant ST/S000321/1.
This article profited from discussions during the meetings of the ISSI International Team Magnetic Helicity in Astrophysical Plasmas.
\end{acknowledgements}

%%%%%%%%%%%%%%%%%%%%%%%%
% Bibliography
% Using BibTeX
% \bibliography{/work/lib/latex/solar}
% \bibliography{solar}
% \input{H_partition.bbl}
% \bibliography{solar}
 \bibliographystyle{aa} %% Alternative style: no title, no concluding page
%Copy from .bib file

%%%%%%%%%%%%%%%%%%%%%%%%%%%%%%%%%%%%%%%%%%%%%%%%
\appendix
\section{Derivation of the non-additive terms}\label{s:corrections}
To allow for cancellations between terms, we split contributions from $\surfa$ (respectively, $\surfb$)  into contributions from the interface $\surfS^a$ (respectively, $\surfS^b$) and contributions from the remaining, non-interface boundaries $\surfZ^a$ (respectively, $\surfZ^b$), see \sect{volumes}.

%%%%%%%%%%%%%%%%%%%%
\subsection{Derivation of the $\delta\HH$ term}\label{s:dH_derivation}
Let us start from \eq{dH}
\BE
\dHH = \intv\vA\cdot\vB \, \dV -\intva\vAa\cdot\vB \, \dV -\intvb\vAb\cdot\vB  \, \dV \, . \label{eq:dH_int}
\EE
The solenoidal condition imposes that the normal component of $\vB$ is continuous across $\surfS$ \citep{Berger1984a}. 
However, here we make the stronger assumption, reasonable in applications, that the vector potential $\vA$ is continuous with its derivatives at the interface such that the magnetic field  $\vB$ is continuous there (\ie\  $\vA|_\surfS \in \rm{C}^1$ and, hence, $\vB|_\surfS \in \rm{C}^0$, at least). 

First, we note that
\BE
\left \{ \begin{array}{l l }
           \vB=\Nabla\times\vA = \Nabla\times\vAa  & \quad \forall \ \vx \in \vola \\
           \vB=\Nabla\times\vA = \Nabla\times\vAb  & \quad \forall \ \vx \in \volb \, .\\
         \end{array} \right .    
\label{eq:AaAb}
\EE
Since, \eg\ both $\vA$ and $\vAa$ produce the same field in $\vola$, then from \eq{AaAb} it follows that they can differ from $\vA$ at most by the gradient of a scalar function, \ie
\BE
    \left \{ \begin{array}{l l }
                \vA= \vAa+\Nabla \chi^a  & \quad \forall \ \vx \in \vola \\
                \vA= \vAb+\Nabla \chi^b  & \quad \forall \ \vx \in \volb \, . \\
             \end{array} \right . 
\label{eq:Aab}
\EE
Using the continuity of $\vA$ and, hence of the RHSs of the above equations, we can split the first integral in \eq{dH_int} into the sum of the integrals on $\vola$ and $\volb$ to obtain 
\BE
 \dHH = \intva \Nabla \chi^a\cdot\vB \, \dV + \intvb \Nabla \chi^b\cdot\vB \, \dV \, ,
\EE
and, by means of  Gauss theorem and the solenoidal condition for $\vB$, 
\BE
 \dHH = \intsa \chi^a \left (\vB \cdot \dSa \right) + \intsb \chi^b \left (\vB \cdot \dSb \right) \, .
\EE
We now split the surface integrals into interface and non-interface contributions obtaining
\BE
 \dHH = \ints \chi^{ab} \left (\vB \cdot \dS \right) 
      + \int_\surfSa \left ( \chi^a -\chi^b \right) \vB \cdot \dSa \, , \label{eq:dH_fin}
\EE
where we introduced the notation of \eq{Zab} for the gauge functions $\chi^a$ and $\chi^b$, and we used \eq{split_v} in the first integral on the RHS, and that $\hatnb =-\hatna$ on $\surfSb$ in the second. 
We anticipate that the first term on the RHS of \eq{dH_fin} cancels with the homologous term from $\dHHmix$  by virtue of \eq{Bp_bc}, and it is therefore omitted from \eq{dH_f}. 
Then, \eq{dH_fin} implies \eq{dH_f} with $\chi = \chi^a -\chi^b$.

%%%%%%%%%%%%%%%%%%%%
\subsection{Derivation of the $\DHp$ term}\label{s:dHp_derivation}
The explicit form of \eq{dHp} is
\BE
\dHHp = \intv\vAp\cdot\vBp \, \dV -\intva\vApa\cdot\vBpa \, \dV -\intvb\vApb\cdot\vBpb  \, \dV \, . \label{eq:dHp_int}
\EE
This cannot be computed as $\dHH$ above because, in general, $\vBpa$ and $\vBpb$ are different from $\vBp$ in the volume $\vola$ and $\volb$, respectively.  
Below, we rather consider explicitly that the reference fields are all potential in their respective domains and must satisfy the gauge-invariance conditions  \eq{Bp_bc},  \ie{}  they satisfy
\BE
\left \{ \begin{array}{l l }
 \vBp = \Nabla \phi  \\
 \hatn \cdot \Nabla \phi|_\surf = \hatn \cdot \vB |_\surf \, ,
 \end{array} \right . 
 \label{eq:Bp_def}
\EE
 for the reference field in $\vol$, and 
 \BE
 \left \{ \begin{array}{l l }
  \vBpa = \Nabla \phi^a  \\
  \hatna \cdot \Nabla \phi^a|_\surfa = \hatn^a \cdot \vB |_\surfa
 \end{array} \right .
 \text{and }
 \left \{ \begin{array}{l l }
  \vBpb = \Nabla \phi^b  \\
  \hatnb \cdot \Nabla \phi^b|_\surfb = \hatn^b \cdot \vB |_\surfb \, ,
 \end{array} \right .
 \label{eq:Bpab_def}
 \EE
 for the reference fields in $\vola$ and $\volb$, respectively.

Let us now use \eq{Bp_def} and \eq{Bpab_def} and the Gauss theorem in \eq{dHp} to readily derive
\BE
\DHp=\DHp^{Coul} + \DHp^{Surf} \, ,
\EE
where
\BA
\DHp^{Coul} =
      &+& \intva \phi^a \left (\Nabla \cdot \vApa \right ) \dV + \intvb \phi^b \left (\Nabla \cdot \vApb \right ) \dV  
      \nonumber \\
      &-& \intv \phi \left (\Nabla \cdot \vAp \right ) \dV   
      \label{dHp_coul} \, , \\
 \DHp^{Surf}=
       &-& \intsa \phi^a \left ( \vApa\cdot\dSa \right) -\intsb \phi^b \left ( \vApb\cdot\dSb \right)
       \nonumber \\
       &+& \ints \phi \left ( \vAp\cdot\dS \right)  
       \label{dHp_surf} \, . 
\EA
The surface term $\DHp^{Surf}$ can be further reorganized by spitting it into interface and non-interface contributions to obtain
\BA
\DHp^{Surf}  &=& \ints  \left ( \phi \vAp -\phi^{ab}\vAp^{ab} \right ) \cdot \dS  \nonumber \\
             &-& \intsS \left ( \phi^a\vApa -\phi^b\vApb \right ) \cdot \dSa
      \label{dHp_surf_1} \, , 
\EA
where, in the last term,  we have used that $\hatna=-\hatnb$ on $\surfS$ and the notation of \eq{Zab} for the scalar potentials $\phi^a$ and $\phi^b$ and  the vector potentials of the potential fields $\vApa$ and $\vApb$.

%%%%%%%%%%%%%%%%%%%%%%%%%%%%%%%%%%%%%%%%
\subsection{Derivation of the $\delta\HH_{\rm mix}$ term}\label{s:dHmix_derivation}
$\delta\HH_{\rm mix}$, \Eq{dHmix}, can be written in terms of solely surface integrals using \eq{HmixS} as
\BA
 \dHHmix &=& \ints \left ( \vA \times \vAp \right) \cdot \dS - 
             \intsa\left ( \vAa\times \vApa\right) \cdot \dSa \nonumber \\
         &-& \intsb\left ( \vAb\times \vApb\right) \cdot \dSb \, .
\label{eq:dHmix_1}
\EA
Using the continuity of $\vA$ and $\vAp$ across $\surfS$ and the definitions of \eqss{split_a}{split_v}, we can split the first integral on $\surf$ into the sum over $\surfa$ and $\surfb$ by adding and subtracting the interface contributions as 
\BE
\ints = \quad \intza + \intzb  = \quad \intsa + \intsb - \intsSa - \intsSb  \, ,
\EE
and, using \eq{Aab} to eliminate the vector potential $\vA$, we have
\BA
&\phantom{+}&  \ints \left ( \vA \times\vAp \right) \cdot \dS =                                                    \nonumber \\
&\phantom{+}&  \intsa  \left ( \vAa \times \vAp \right) \cdot \dSa + \intsb  \left ( \vAb \times \vAp \right) \cdot \dSb    \nonumber \\ 
&+& \intsa  \left ( \Nabla\chi^a \times \vAp \right) \cdot \dSa + \intsb  \left ( \Nabla\chi^b \times \vAp \right) \cdot \dSb \nonumber \\
&-& \intsSa \left ( \left ( \vAa \times \vAp \right)  -  \left ( \vAb \times \vAp \right) \right )\cdot \dSa                \nonumber \\
&-& \intsSa \left ( \left ( \Nabla\chi^a - \Nabla\chi^b \right ) \times \vAp  \right ) \cdot \dSa   \, .
\label{eq:dHmix_2}
\EA
The identity
\BE
 \Nabla \times \left ( \vAp\ \chi^{a} \right ) = \chi^{a} \vBp + \Nabla \chi^{a} \times \vAp  \, ,
 \label{eq:vect_id}
\EE
in $\surfa$, and analogous expression for $\vAp\ \chi^{b}$ in $\surfb$, where all vector fields satisfy the necessary continuity conditions, can be now used to re-write the second line in the RHS of \eq{dHmix_2} as
\BA
&&\intsa  \Nabla \times \left (\chi^a \vAp \right) \cdot \dSa + \intsb  \Nabla \times \left (\chi^b \vAp \right) \cdot \dSb \nonumber \\
&-& \intsa \chi^a \left (\vBp\cdot \dSa \right ) - \intsb \chi^b \left (\vBp\cdot \dSb \right )  \, ,
\label{eq:int_vect_id}
\EA
where the first two terms are identically zero because the curl of any (sufficiently continuous) vector field is solenoidal, and the flux trough a closed surface of a solenoidal field vanishes. 
Substituting back in \eq{dHmix_1},
\BA
\dHHmix &=& \intsa  \left ( \vAa \times \vAp \right) \cdot \dSa + \intsb  \left ( \vAb \times \vAp \right) \cdot \dSb    \nonumber \\
        &-& \intsa \chi^a \left (\vBp\cdot \dSa \right ) - \intsb \chi^b \left (\vBp\cdot \dSb \right )                              \nonumber \\
        &-& \intsSa \left ( \left ( \vAa \times \vAp \right)  -  \left ( \vAb \times \vAp \right) \right )\cdot \dSa                \nonumber \\
        &-& \intsSa \left ( \left ( \Nabla\chi^a - \Nabla\chi^b \right ) \times \vAp  \right ) \cdot \dSa                           \nonumber \\
        &-& \intsa  \left ( \vAa\times \vApa\right) \cdot \dSa - \intsb\left ( \vAb\times \vApb\right) \cdot \dSb  \, .
\label{eq:dHmix_3}
\EA
Splitting into interface and non-interface contributions we get
\BA
\dHHmix 
        &=& \intza  \left ( \vAa \times \left ( \vAp - \vApa \right) \right) \cdot \dSa  + \intzb  \left ( \vAb \times \left( \vAp -\vApb \right )\right) \cdot \dSb    \nonumber \\
        &-& \intsSa  \left ( \vAa\times \vApa\right) \cdot \dSa - \intsSb\left ( \vAb\times \vApb\right) \cdot \dSb       \nonumber \\
        &-& \intsSa \left (\chi^a-\chi^b \right ) \left (\vBp\cdot \dSa \right )                                          \nonumber  \\
        &-& \intsSa \left ( \left ( \Nabla\chi^a - \Nabla\chi^b \right ) \times \vAp  \right ) \cdot \dSa                 \nonumber \\
        &-& \ints \chi^{ab}\left (\vBp\cdot \dS \right )    \, ,                   
\label{eq:dHmix_4}
\EA
where the continuity of $\vBp$ across $\surfS$ was used in the third line of the RHS and the notation of \eq{Zab} for the gauge functions $\chi^a$ and $\chi^b$ in the last one.
Using the notation in  \eq{Zab} also for the  vector potentials $(\vAa,\vAb)$ and $(\vApa, \vApb)$, and the definition of \eq{split_v}, we can formally write 
\BA
\dHHmix
        &=& \ints  \left ( \vA^{ab} \times \left ( \vAp - \vAp^{ab}\right) \right) \cdot \dS   \nonumber \\
        &-& \intsSa\left ( \left ( \vAa\times \vApa\right) - \left ( \vAb\times \vApb\right) \right )\cdot \dSa      \nonumber \\
        &-& \intsSa \left (\chi^a-\chi^b \right ) \left (\vBp\cdot \dSa \right )                                          \nonumber  \\
        &+& \intsSa \left ( \left ( \vAa-\vA^b  \right ) \times \vAp  \right ) \cdot \dSa                 \nonumber \\
        &-& \ints \chi^{ab}\left (\vBp\cdot \dS \right )   \, ,                    
\label{eq:dHmix_5}
\EA
where we used that, on $\surfSa$, is 
\BE
 \Nabla \chi=\Nabla \chi^a-\nabla \chi^b = - (\vAa -\vAb) \, ,
 \label{eq:chi_ab_sigma}
\EE
by virtue of \eq{Aab} and the continuity of $\vA$ across  $\surfS$, with $\chi=\chi^a-\chi^b$.
Note that, while $\chi^a$ and $\chi^b$ are defined in the entire subvolumes $\vola$ and $\volb$, respectively, the gauge function $\chi$ is defined only on the interface $\surfS$ and is a function of the interface variables only.

The last term on the RHS of \eq{dHmix_5}  cancels with the homologous term in \eq{dH_fin} by virtue of \eq{Bp_bc}, and it is therefore omitted from \eq{dHmix_f}.
%%%%%%
\section{Gauge-invariance of the additivity formula}\label{s:gi-additivity}
In this section the invariance of \eq{dHv} with respect to gauge transformations of the vector potentials is proven. 
First notice that each of \eqss{dH_f}{dHmix_f} are invariant if we interchange $a \leftrightarrow b$, since $\dSa = - \dSb$ on $\surfS$. 
Hence, it suffices to check gauge invariance under gauge changes of $\vA$, $\vAp$ $\vAa$, and $\vApa$. 
We consider these gauge transformations in turn:

1. $\vA \rightarrow  \vA + \Nabla \psi$. 
Notice that $\vA$ does not explicitly appear in any of the expressions \eqss{dH_f}{dHmix_f}.
However, since we are not changing $\vAa$ or $\vAb$, transforming  $\vA \rightarrow  \vA + \Nabla \psi$ corresponds to the change $\chia \rightarrow (\chia +\psi)$  and $\chib \rightarrow (\chib+\psi)$, as follows from \eq{Aab}.
Since nevertheless these potentials appear only in the combination $\chib-\chia$ then  $\delta \Hv$ is invariant with respect to the transformation.

2. $\vAp \rightarrow  \vAp + \Nabla \psi$. 
\Eq{dH_f} clearly shows that $\dHH$ is unchanged by this transformation. 
From \eq{dHp_f} we have for $\dHHp$ that
\BE
\dHHp \rightarrow  \dHHp  - \intv \phi \Delta \psi \dV + \ints \phi \Nabla \psi \cdot \dS \, , 
\label{eq:dHHp_Ap_1}
\EE
Using Gauss theorem twice and \eq{Bp_def} we can write the first integral as 
\BE
\intv \phi \Delta \psi \dV = \ints \phi \Nabla \psi \cdot \dS - \ints \psi \vBp \cdot \dS  + \intv \psi \left ( \Nabla \cdot \vBp \right )\dV \, ,  \nonumber
\EE  
where the last term vanishes since $\vBp$ is solenoidal.
Substituting in \eq{dHHp_Ap_1} we have 
\BE
\dHHp \rightarrow  \dHHp + \ints \psi \vBp \cdot \dS \, .
\label{eq:dHHp_Ap_2}
\EE
Using \eq{dHmix_f} and \eqs{split_a}{split_b}, the gauge transformation of $\dHHmix$ is 
\BE
\dHHmix \rightarrow  \dHHmix  + \intsa \vAa \times \Nabla \psi \cdot \dSa +  \intsb \vAb \times \Nabla \psi \cdot \dSb  \, , 
\label{eq:dHHmix_Ap_1}
\EE
where the first integral can be written, using \eq{AaAb}, as 
\BE
\intsa \vAa \times \Nabla \psi \cdot \dSa = \intsa \Nabla \times \left(\psi\vAa \right ) \cdot \dSa - \intsa \psi \vB  \cdot \dSa  \, , 
\label{eq:dHHmix_Ap_2}
\EE
where the first integral on the RHS vanishes. 
A similar expression can be derived for  the second integral in \eq{dHHmix_Ap_2}, and, substituting back, we have 
\BA
\dHHmix &\rightarrow&  \dHHmix  -  \intsa \psi \vB  \cdot \dSa -  \intsb \psi \vB  \cdot \dSb \nonumber \\
        &=& \dHHmix - \ints \psi \vB \cdot \dS \, ,
\label{eq:dHHmix_Ap_3}
\EA
where, in the second line, we have used again \eqs{split_a}{split_b} to separate the interface from the non-interface contributions, and $\dSa=-\dSb$.
Hence, considering \eq{dHHp_Ap_2}, \eq{dHHmix_Ap_3}, and \eq{Nabla_phi_bc} we have that overall $\delta \Hv$ is unchanged by this transformation.

3. $\vAa \rightarrow  \vAa + \Nabla \psi$.
Since we are not changing $\vA$, it follows from \eq{Aab} that we must transform  $\chia \rightarrow (\chia -\psi)$.
Therefore, from \Eq{dH_f}, we have
\BE
\dHH \rightarrow  \dHH  +  \intsS \psi \vB \cdot \dSa  \, .
\label{eq:dHH_Aa_1}
\EE
$\vAa$ does not appear in \eq{dHp_f}, hence  $\dHHp$ is unchanged by this transformation. 
On the other hand, using \eq{split_a} we have that \eq{dHmix_f}  transforms as 
\BE
\dHHmix \rightarrow  \dHHmix  + \intsa \Nabla \psi \times \left (\vAp -\vApa \right ) \cdot \dSa - \intsS \psi \vBp \cdot \dSa \, ;
\nonumber
\EE
with analogous manipulation as in \eq{dHHmix_Ap_2}, the first integral can be rearranged as 
\BA
\intsa \Nabla \psi \times \left (\vAp -\vApa \right ) \cdot \dSa &=& \intsa \psi  \left (\vBp -\vBpa \right ) \cdot \dSa \nonumber \\
              &=& \intsS \psi \left (\vBp -\vBpa \right ) \cdot \dSa  \, ,
\EA
where the non-interface contribution vanishes because $\hatn\cdot\vBp$ and  $\hatn\cdot\vBpa$ are the same there, by virtue of \eq{Nabla_phi_bc} and \eq{Nabla_phia_bc}.
It follows that 
\BA
\dHHmix &\rightarrow&  \dHHmix  + \intsS \psi \left (\vBp -\vBpa \right ) \cdot \dSa -\intsS \psi \vBp \cdot \dSa \nonumber \\
        &=&            \dHHmix -\intsS \psi \vBpa \cdot \dSa \, ,
\label{eq:dHH_Aa_2}
\EA
and therefore, considering \eqs{dHH_Aa_1}{dHH_Aa_2} and \eq{Nabla_phia_bc}, we have that overall $\delta \Hv$ is unchanged by this transformation.

4. $\vApa \rightarrow  \vApa + \Nabla \psi$.
Also in this case  $\dHH$ is unchanged by this transformation. 
From \eq{dHp_f} we have for $\dHHp$ that
\BE
\dHHp \rightarrow  \dHHp  + \intva \phia \Delta \psi \dVa - \intsa \phia \Nabla \psi \cdot \dSa \, , 
\label{eq:dHHp_Apa_1}
\EE
which is similar to \eq{dHHp_Ap_1} but written for $(\phia,\vola)$ rather than $(\phi,\vol)$, and with opposite signs of the integrals.
With similar transformations as in \eqss{dHHp_Ap_1}{dHHp_Ap_2}  we find
\BE
\dHHp \rightarrow  \dHHp - \intsa \psi \vBpa \cdot \dSa \, .
\label{eq:dHHp_Apa_2}
\EE
For  $\dHHmix$, the gauge change implies the transformation
\BE
\dHHmix \rightarrow  \dHHmix  - \intsa \vAa \times \Nabla \psi \cdot \dSa   \, ,
\label{eq:dHHmix_Apa_3}
\EE
where \eq{split_a}  was used. 
Using \eq{dHHmix_Ap_2} we have 
\BA
\dHHmix &\rightarrow&  \dHHmix  + \ints \psi \vB \cdot \dSa  \nonumber \\
        &=& \dHHmix  + \ints \psi \vBpa \cdot \dSa \, , 
\label{eq:dHHmix_Apia_4}
\EA
where \eq{Nabla_phia_bc} was used to obtain the second line. 
Once again, $\delta \Hv$ is unchanged overall.

This completes the proof.
%%%%%%%%%%%%%%%%%%%%%%%%%%%%%%%%%%%%%%%%%%%%%%%%%%%%%%%%%%%
%
\end{document}